\documentclass[journal]{IEEEtran}

\usepackage{amssymb}
\usepackage{latexsym}

\usepackage{url}
\usepackage{xcolor}
\usepackage{adjustbox,multirow}
\usepackage{array,booktabs}
\definecolor{newcolor}{rgb}{.8,.349,.1}

\usepackage{tikz}
\usepackage{tkz-graph}
\usepackage{subfigure}
\def\newblock{}

\usetikzlibrary{arrows,decorations.pathmorphing}
\usetikzlibrary{shapes,snakes}
\usetikzlibrary{plotmarks}

\usetikzlibrary{shadows,  backgrounds}
\usepackage{pgf}
\newcolumntype{R}[2]{    >{\adjustbox{angle=#1,lap=\width-(#2)}\bgroup}    c    <{\egroup}}
\newcommand*\rot{\rotatebox{90}}
\newcommand{\ar}[1]{{\color{black}#1}}
\newcommand{\modif}[1]{{\color{black}#1}}
\newcommand{\modifR}[1]{{\color{black}#1}}

\begin{document}

\title{Histogram of Gradients of Time-Frequency Representations for
Audio Scene Detection}

\author{ A. Rakotomamonjy, G. Gasso 
 \thanks{This work has been partly supported by the French ANR (12-BS02-004).}
\thanks{
 \newline AR is with LITIS EA 4108, Universit\'e de Rouen, France. alain.rakoto@insa-rouen.fr
 \newline GG is with LITIS EA 4108, INSA de Rouen, France. gilles.gasso@insa-rouen.fr }
}

\maketitle

\begin{abstract}
This paper addresses the problem of audio scenes classification
and contributes to the state of the art by proposing a novel feature. We build this feature by considering
histogram  of gradients (HOG) of an audio scene time-frequency representation.
Contrarily to classical audio features like MFCC, we make the hypothesis
that {histograms} of gradients are able to encode some relevant
informations in a time-frequency {representation:} namely, the local
direction of variation (in time and frequency) of the signal spectral power.
In addition, in order to gain more invariance and robustness, 
{histograms} of gradients are locally pooled. 
We have evaluated the relevance of {the novel feature} by comparing its performances
with state-of-the-art competitors, on several datasets, including a novel
one that we provide, as part of our contribution. This dataset, that we
make publicly available, involves $19$ classes and contains about $1500$
minutes of audio scene recordings. We thus believe that it may be the next standard dataset for evaluating audio scene classification algorithms. 
Our comparison results clearly show that the HOG-based features outperform
its competitors.
\end{abstract}


\section{Introduction}
The problem of recognizing acoustic environments is known
as the problem of audio scene classification and it
is one of the most difficult task in the general context
of computational auditory scene analysis (CASA) \cite{wang06:_comput}. 
This classification task is of primary importance
in the domain of machine listening since it
is strongly related to the context in which the
acquisition device capturing the audio scene lives. 
Typically, in order to get some context awareness, a machine, say a smart-phone or any {mobile} electronic device, should be able to predict the 
environment in which it currently resides. {The main goal is to} help the
machine {adapting} itself to the context of the user (for instance
by automatically turning off the ring tone in some situations).
Such awareness can be  brought {through} vision or audio scene analysis. While most of the efforts have focused on vision, there is now a growing interest
of {environment} recognition based on audio modality.

Audio scene classification is a
very complex problem since a recording related to a given location
can be potentially composed of a very large amount of single
sound events while only few of these events provide some information on the 
scene of the recording.  
More specifically, an audio scene is associated to
a recording taken at a given location and this location is expected 
to generate some acoustic events that make
it distinguishable from other audio scenes. These
discriminative acoustic events may be produced by different phenomena
and  they may have a very large variability.

Hence, the recent works on audio scene classification have 
devoted much efforts on designing methods and algorithms
for automatically extracting audio features that
capture the specificities of these events. 
The natural hope is that the designed features
are still able to  capture the discriminative
power of a given audio event.

For instance, following its success in speech recognition,
one of the most prominent features that has been considered
for audio scene recognition are mel-frequency cepstral 
coefficients (MFCC) \cite{aucouturier07,ma06:_acous_envir_class,cauchi11:_non,hu12:_combin}. 
These features are typically used in conjunction with different
machine learning techniques in order to capture the variations
that help in discriminating scenes. For instance, 
\cite{aucouturier07} consider a Gaussian Mixture Model
for estimating the distribution of the MFCC coefficients,
while \cite{lee13:_acous} have proposed a sparse feature
learning approach for capturing relevant MFCC coefficients.

In addition to MFCC, several {kinds} of features  have also been
evaluated for solving this problem of audio scene recognition.
 \cite{chu09:_envir_sound} proposed an ensemble of  time-frequency features
obtained from a matching pursuit decomposition of the audio
signal. Recently, \cite{geiger13:_large_svm} have considered a  large
set of features, including spectral, energy-based and voicing-related
features. Another family of relevant features
can be obtained from MFCC by considering recursive
quantitative analyzing (RQA) as introduced by \cite{roma13:_recur}. 
RQA has the advantage to allow the analysis of recurrent
behaviour in the MFCC coefficients over time. 
According to the recent D-Case challenge on audio
scene recognition \cite{giannoulis13:_detec}, combining MFCC features with RQA features extracted from MFCC yield to an highly 
efficient set of features.

Another trend aims at building higher-level features from
the time-frequency representations of the audio scene.
In this context, \cite{cotton11:_spect} have investigated
methods for automatically extracting spatio-temporal  patches 
that are discriminative of the audio scene. Typically, these patches are obtained through a non-negative matrix
factorization of a time-frequency representation. \cite{benetos12:_charac} have followed similar ideas but instead of considering
NMF they employed a probabilistic model denoted as probabilistic
latent component analysis. Likewise, other works 
propose features, like texture-based features that are directly computed from  time-frequency representations \cite{Yu:_Audio2009,dennis13:_image_featur}. 

In this paper, we follow this trend and propose a novel feature for automatic
recognition of audio scene. The main originality of
the proposed feature is the use of histogram of
gradients (HOG) on time-frequency representations. 
These HOG features have been genuinely introduced for
human detection in images  \cite{dalal2005histograms}
but we strongly believe that their properties make them highly
valuable for extracting relevant features based on time-frequency representation (TFR).
Indeed, while  MFCC can also {be} considered
as features extracted from a time-frequency  {(TF)} representation,
they essentially capture non-linear information on the power
spectrum of the signal. 
Instead, histogram of gradients of a TF representation provides
information on the  spectral power direction of variation. 
For instance, if an audio scene has been obtained in 
{a} bus that it is accelerating or decelerating, we expect 
the \emph{chirp} effect present in the TFR to be 
captured and better discriminated by the histogram of gradients representation,
than by MFCC.
This property will be empirically illustrated in the sequel
and it provides rationale that for audio scene recognition the
novel feature we present is strongly relevant.

Algorithms for audio scene recognition have to be validated
and evaluated on some datasets. In order to make  comparisons
of different {designs of features including signal processing set-ups or different learning techniques} possible, these datasets should be 
publicly available. 
The recent D-Case challenge \cite{giannoulis13:_detec} is  an excellent initiative
of this kind although its number of examples is  limited ($100$ for
the publicly available examples). 
Hence, another contribution we present is a new dataset
for audio scene recognition. It is based on  about $1500$ minutes of recording
on different  locations (up to $19$). This dataset is
publicly available and we expect that it will become one of the standard
benchmarks for audio scene recognition.

The paper is organized as follows: we first describe the pipeline
we propose for extracting our novel HOG-based feature. Then, as one of our
contribution is also to introduce a novel benchmark dataset, we carefully
 {detail}  all the datasets we considered for evaluating our feature
and its competitors as well as the experimental protocol
we employed for the comparisons. Extensive experimental analyses have been carried out and they show that our HOG-based feature achieves state-of-the-art performances on all datasets.  As we advocate result reproducibility,
 all the codes used for this work will  be made publicly available
on the author's website.

\section{From signal to histogram of gradients features}
This section describes the feature extraction pipeline
we propose for analyzing audio scene signals. We first
provide the big picture before detailing each part of
the flowchart.

\subsection{The global feature extraction  scheme}

The features we propose for recognizing audio scenes are based
on some specific information extracted from a time-frequency
{representation} (TFR) of the signal. After the TFR image has been
computed, it is processed so as to attenuate some spurious
noises that may hinder  relevant
information related to  high-energy time-frequency structures. 
Afterwards, the resulting processed time-frequency representation
image is used as input of our histogram of {gradients} feature extraction.
In a nutshell, the idea of histogram of {gradients} is to locally analyze
the direction of energy's variation in the time-frequency representation.
As detailed in the sequel, the local HOG {informations over the whole TF image} are combined in order to generate
the final feature {vector}. The dimension of this vector depends
on the number of bin in the (local) histogram and on how
all the local histograms are pooled together in order to 
form the final feature vector. 

The block diagram of this feature extraction scheme is illustrated
in Figure \ref{fig:block}. \ar{Each block of this diagram is
discussed in the next paragraphs.}

\pgfdeclarelayer{background}
\pgfdeclarelayer{foreground}
\pgfsetlayers{background,main,foreground}

\definecolor{chaptercolor}{RGB}{255,254,255}
\definecolor{chaptercolortwo}{RGB}{104,104,255}
\colorlet{fondun}{blue!20!white!20!gray}
\colorlet{fonddeux}{red!50!blue!50!gray}

\definecolor{colorsection}{RGB}{112,70,109}

\definecolor{colorsubsection}{rgb}{0.2,0.4,0.65}
\colorlet{colorsubsubsection}{green!50!blue!20!gray}
\tikzstyle{block} = [draw= colorsubsection, drop shadow, line width= 0.05cm, rectangle, minimum height=2.5em, minimum width=10em, rounded corners, top color= white, bottom color	= chaptercolortwo!20]
\tikzstyle{blocktwo} = [draw= colorsubsection, line width= 0.05cm, drop shadow,  rectangle, minimum height=2em, minimum width=8em, rounded corners, top color= white, bottom color	= chaptercolortwo!20]
\tikzstyle{blockthree} = [draw= red!50!black!50, line width= 0.05cm, drop shadow,  rectangle, minimum height=2em, minimum width=8em, rounded corners, top color= white, bottom color	= red!50!black!20]

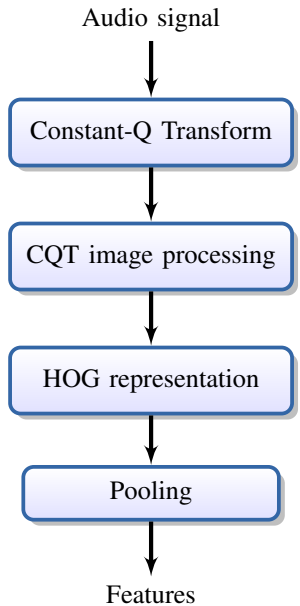
\begin{figure}[t]
\begin{center}
  \begin{tikzpicture}[auto, scale = 1,>=latex']
        \node (Sig) { Audio signal};
    \node [block, below of=Sig, node distance = 1.5cm] (CQT) {
    \begin{minipage}{10em}
	\centering
	Constant-Q Transform
    \end{minipage}
    };
    \draw[ultra thick, ->] (Sig) -- (CQT);
    \node [block, below of=CQT, node distance = 1.65cm] (filter) {
    \begin{minipage}{10em}
	\centering
	CQT image {processing}
    \end{minipage}
    };
    \draw[ultra thick, ->] (CQT) -- (filter);
    \node [block, below of=filter, node distance = 1.65cm] (hog) {
    \begin{minipage}{10em}
	\centering
	HOG  representation
    \end{minipage}
    };
    \draw[ultra thick, ->] (filter) -- (hog);
    \node [blocktwo, below of=hog, node distance = 1.5cm] (pool) {
    \begin{minipage}{9em}
	\centering
	Pooling
    \end{minipage}
    };
    \draw[ultra thick, ->] (hog) -- (pool);
    \node [below of=pool, node distance = 1.35cm] (features) {
    \begin{minipage}{8em}
  	\centering
     Features
    \end{minipage}
   };
      \draw[ultra thick, ->] (pool) -- (features);
\end{tikzpicture}
\end{center}
\caption{
 Block Diagram of the HOG-based feature extraction 
\label{fig:block}}
\end{figure}

\subsection{From signals to TFR images}

Because of their non-stationary nature, sounds are typically represented
on a short-time power frequency representation, which idea is to capture
the power spectrum of the signal on a varying short local window. 
A large part of the literature on sound recognitizion problems use
such  time-frequency representations of sound. Depending on the task
at hand, they either consider wavelet-based \cite{neumann05:_effic_wavel_adapt_for_hybrid,lung08:_featur_riesz}, MFCC-based
\cite{cheng10} or short-time Fourier-based representations \cite{cowling03:_compar}.
In this work, we do not depart from this widely adopted framework
and choose to represent the signal according to a
constant-Q transform  \cite{brown91:_calcul_q}. Contrarily
to a short-time Fourier transform, this transform provides
a frequency analysis on a log-scale which makes it more
adapted to sound and music representations \cite{brown91:_calcul_q}.

Once this TFR has been computed, we now have an image that can be
processed as such. {In} order to obtain a processing
system independent of the signal length and sampling frequency, as well as the
CQT parameters, we have chosen to resize all TFRs to a $512 \times 512$
image. \modif{This resizing is performed on the CQT matrix by means
of a bicubic interpolation. Hence, the image we obtain is not exactly
equivalent to a CQT with a total of $512$ frequency bins
but it  preserves the time-frequency structures of the audio scene as one can see in Figure \ref{fig:toychirp}.}

\modif{
Then, depending on  the TFR
images, some image processing tools can be used so as
to enhance relevant time-frequency structures. In our work, because we have few prior knowledge on 
the signal noise, we have just applied a mean filtering so
as to smooth the TFR image.  
Our goal with this smoothing is to reduce strong 
local variations in the image that will tend to produce
high gradients, which may be not relevant for
audio scene recognition. The size of the average kernel  
for mean filtering can be considered then as an hyperparameter
of the feature extraction scheme and its influence will be
investigated in the experimental analysis. 
}

\subsection{Histogram of gradients}

\begin{figure*}[ht]
  \centering
~\hfill
  \includegraphics[width=5cm]{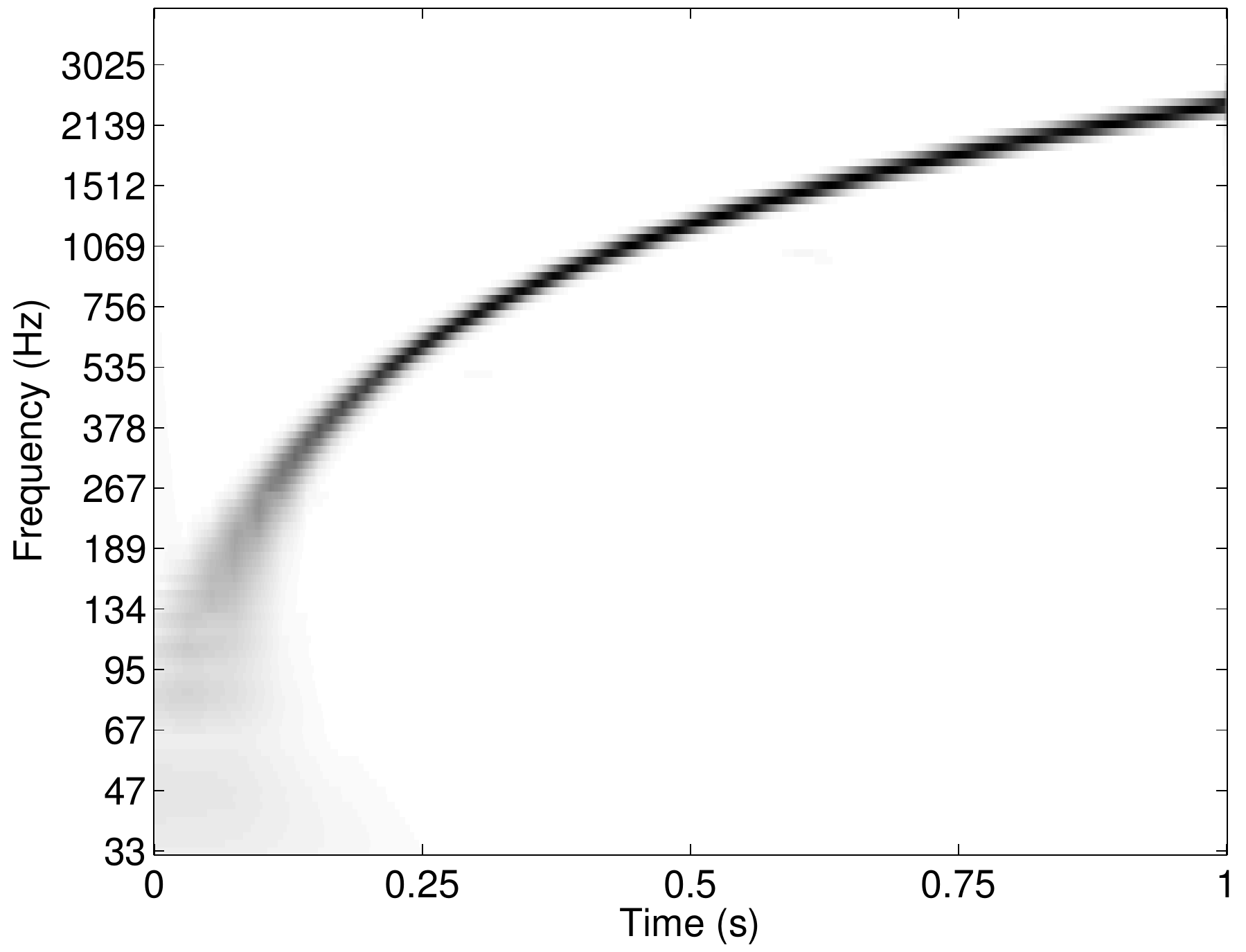}
\hfill
  \includegraphics[width=5cm]{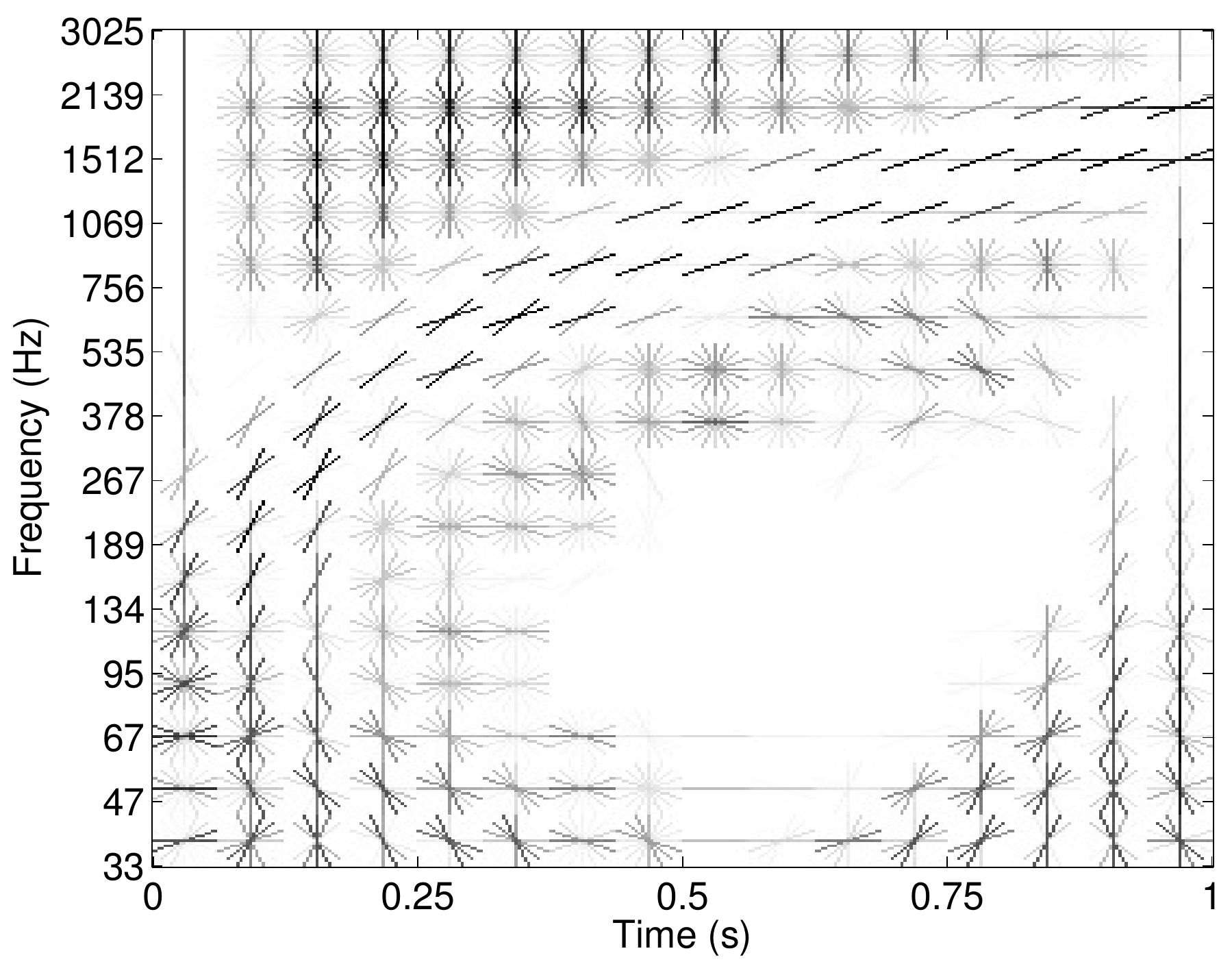}
\hfill~
  \caption{Example of HOG for a toy linear chirp. (left) 
$512 \times 512$ image of the CQ transform of the signal.
(right) histogram of {gradients} {representation} of the signal.
Note that for a sake of interpretation, each cell in the plots
represents the {occurrence} of edge orientation in the cell and the 
darker the orientation is, the more present  the orientation {is}.
We thus note that along the chirp, the HOG representation
correctly captures the direction of energy variation.   }
  \label{fig:toychirp}
\end{figure*}

Histogram of gradients have been originally introduced
by  \cite{dalal2005histograms} for human detection in images. 
Our main objective in this feature extraction stage is to  capture the shape of
some time-frequency structures in the hope that such
structures are relevant for characterizing an audio scene. From works
in computer vision \cite{dalal2005histograms,felzenszwalb10:_objec,minetto13:_t_hog}, it is now well acknowledged that
local shape information can be described through gradient intensity and
orientations. Histograms of gradients basically provide information
about the occurrence of gradient orientations in a localized region
of the images. Hence, they are able to characterize shapes in that regions. 

Two main approaches have been proposed for computing HOG in images \cite{dalal2005histograms,felzenszwalb10:_objec}
and they are both based on the following steps: 
\begin{enumerate}
\item compute the gradient of the TF image
\item compute angles of all pixel gradients
\item split images into non-overlapping cells
\item count the occurrence of gradient orientations in a given cell
\item eventually normalize each cell histogram according to histogram
norm of neighboring cells. 
\end{enumerate}

Variants on this theme are essentially based on whether the gradient
orientations are bidirectional or not, whether the magnitude
of the gradient is taken into account in the counting and on how normalization
factors are computed within block of neighboring cells. For this work, we have
used the implementation in the \emph{VLFeat} toolbox \cite{vedaldi10efficient}.
\modif{\emph{VLFeat} is an open-source computer vision toolbox  that includes
the major functionalities of the most popular computer vision and pattern recognition algorithms. In particular, it implements several histogram-based feature
extraction algorithms including SIFT \cite{lowe2004distinctive} and HOG. 
For more details, we refer the reader to \cite{dalal2005histograms} and \cite{felzenszwalb10:_objec} and to the \emph{VLFeat} Hog tutorial\footnote{http://www.vlfeat.org/overview/hog.html}}.

As an illustration, we depict in Figure \ref{fig:toychirp} the $512 \times 512$
{mean} filtered image of a linear chirp's CQT transform, as well as the resulting histogram of {gradients} we obtain 
for each $32 \times 32 $ cell with $8$ gradient orientations. From the right panel, we remark that the HOG representation properly {captures} the directions of
power spectrum's variation along the high-energy chirp signal. However,
we can also note that several spurious cells depict non-zero histogram
of gradients (top and right bottom parts of the image). They are essentially due to presence of small variations
of gradient in low-energy time-frequency structure  in the {CQ} transform, inducing non-zero gradients. However,
these noisy cells can be easily recognized as having an almost flat histogram,
denoting thus the presence of multiple orientations of gradient in the cells. 
\modif{Note that in our feature extraction scheme, we have not considered
any pre-processing and post-processing strategies for handling these
spurious cells,  they are taken into account as they are into the HOG features.}

After having computed the histogram of {gradients} in the images, we are left
with a representation composed of  histograms in all cells. If we
concatenate all these histograms for yielding the final feature
vector, we obtain a vector whose dimension is large (number of cells $\times$
number of orientations in the histogram). In the
example in Figure \ref{fig:toychirp}, cells are sized $ 32 \times 32$ pixels, this results in vector of dimension $16^{2}  \times 8 = 2048$, $16^{2}$ being the total number of cells.  Of course, this dimensionality
may further increase if we choose to reduce cell's size or increase the number
of orientations in the histogram computation. 
Depending on the number of audio scene examples, it thus may be {beneficial}
to reduce the dimensionality of the problem for instance by pooling {the histograms of gradients}.

\subsection{Time-Frequency Histogram Pooling}

\begin{figure*}[t]
  \centering
~\hfill~
  \includegraphics[width=5cm]{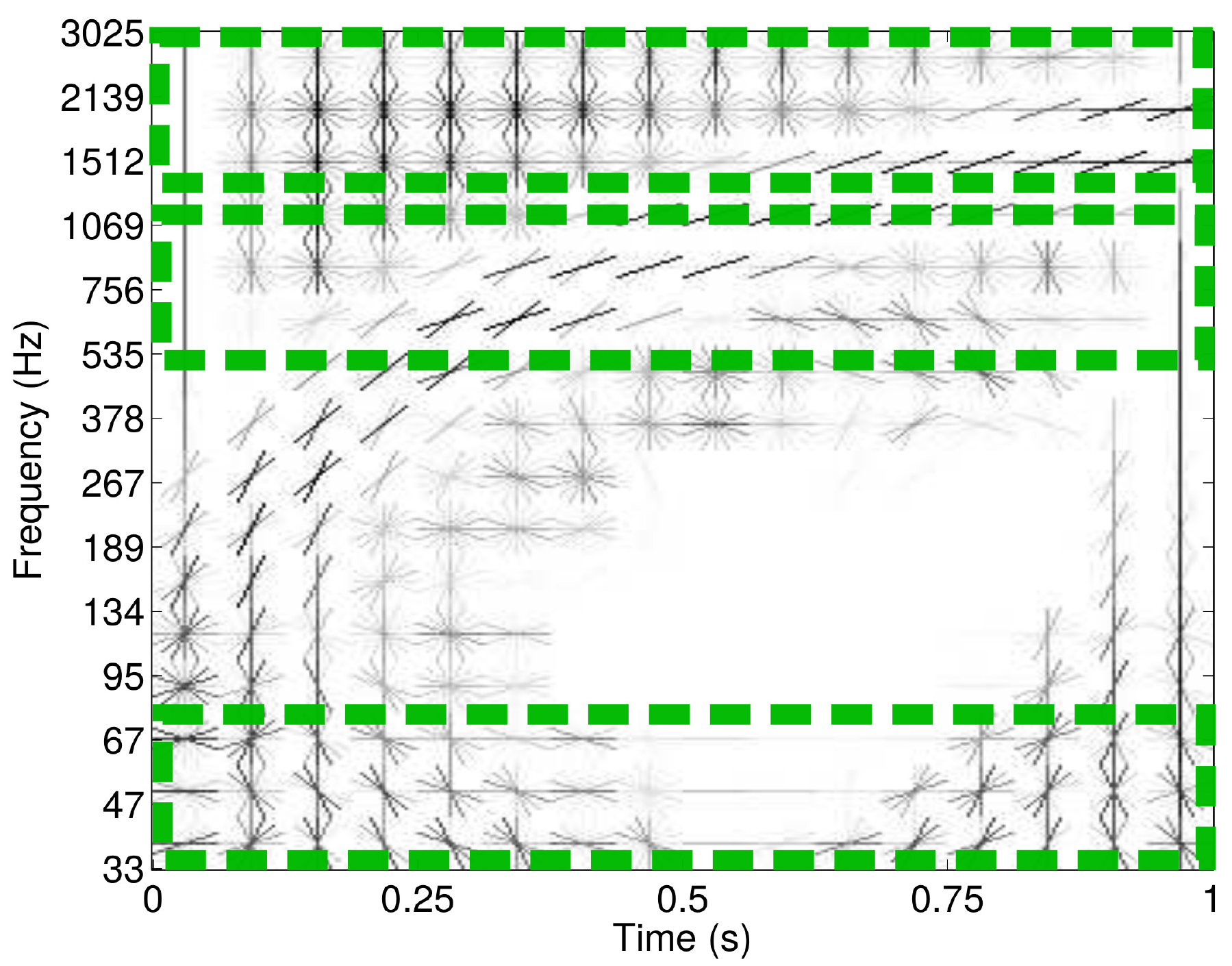}
~\hfill~
  \includegraphics[width=5cm]{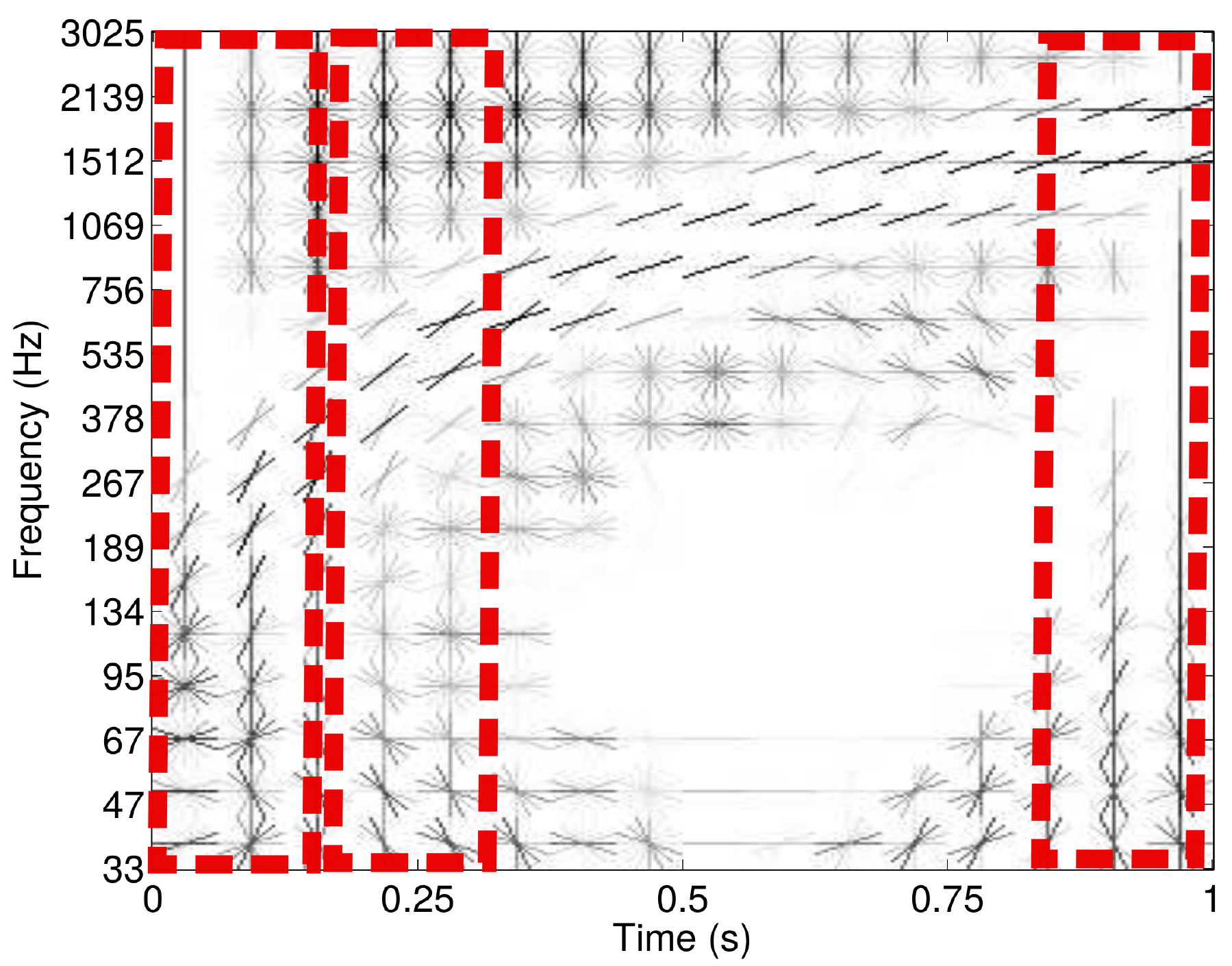}
~\hfill~
  \includegraphics[width=5cm]{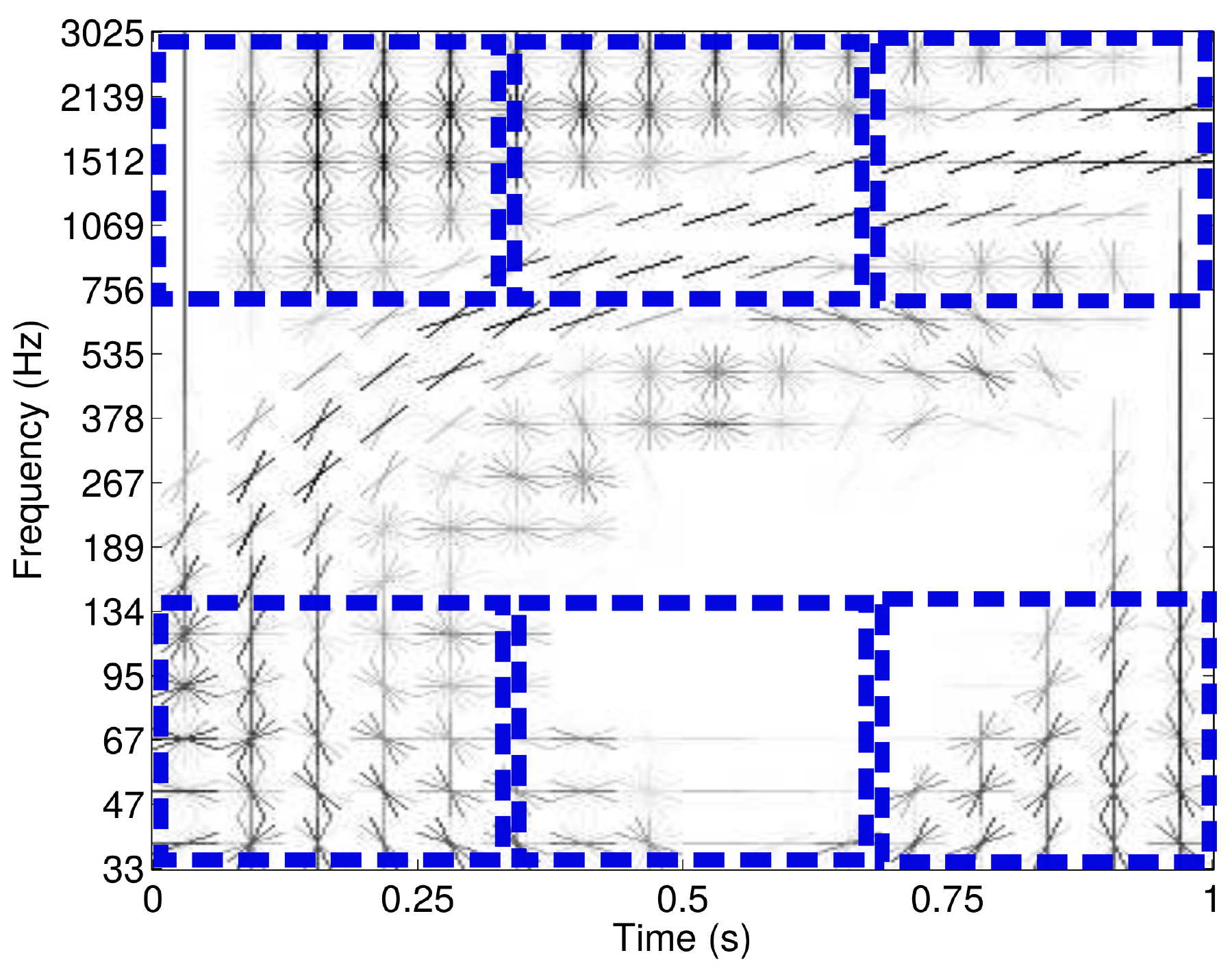}
~\hfill~
  \caption{Illustrating the different types of pooling we
investigate {based on the HOG issued from Figure \ref{fig:toychirp}.} {(Left)} time pooling. {(Middle)} frequency pooling. {(right)}
block-sized pooling. The green, red and blue boxes provide
an example on the regions on which local histograms are averaged. The x-axis
and  y-axis respectively denote the time and frequency axes. Best viewed in color. 
\label{fig:pooling}}
\end{figure*}

Pooling consists in combining the responses of a feature extraction
algorithm computed at nearby locations.  The underlying idea is to summarize local features into another
feature (of lower dimensionality) that is expected to keep relevant information over the {neighborhood}. The pooling  helps in getting a more robust information. This technique is 
a step commonly considered with success in  modern visual recognition algorithms \cite{boureau10}. 
In our case, pooling histograms over neighboring cells aims
at building new histograms that capture information on
time-frequency structures {which} may be larger {than} a cell or that have
been slightly translated in time or frequency. In this work, we will 
investigate several forms of time-frequency region pooling (see Figure \ref{fig:pooling}), while
the pooling operation will be kept fixed as an averaging operator. We will consider the following poolings:
\begin{itemize}
\item Marginalized pooling over time: for this pooling, we average
all histograms along the time axis of the TFR representation. This results
in a feature vector which has lost all temporal information. 
\item Marginalized pooling over frequency: in this case, the averaging is
performed over the frequency axis. Hence, all frequency informations of the HOG
are now merged into a single one.  
\item Block-size pooling: pooling is performed on nearby cells with the size
of the neighorhood being user-defined. 
\end{itemize}

The vector resulting from the concatenation of the pooled histograms
forms now the feature vector that will be used
for learning   the audio scene classifier.

\subsection{Discussions}

Now that we have explained how the HOG on time-frequency representation feature
is obtained,
we want to discuss some properties of these HOG features and their advantages
over features like MFCC for audio scene characterization.

Our initial objective was to design features that is able to characterize
some time-frequency structures that occur in a time-frequency representation.
By construction, since we bin the orientations when counting a given gradient, 
the histogram of gradients is invariant to rotation if this rotation is 
smaller than the bin size. \modif{In our case, rotation would correspond
to a small rotation of a time-frequency structure leading then to
a change of orientation {of} its gradients. Such a situation may
occur for instance in audio scenes capturing a moving object, like
a bus or a car. Indeed,  variations of a bus's acceleration induce
 variations of steepness in the time-frequency structure related to
the sound of that bus. Owing to the binning of the gradient orientation,
the HOG feature will be invariant to these variations. 
 }
Furthermore, as we build an histogram
from a cell of pixels and then average them over a larger region, our
pooled histogram of gradient is invariant to translation over
that region of pooling.

Compared to classical features like MFCC used for audio applications, 
HOG-based features present several benefits. 
For instance, they are, by construction, invariant to small time
and frequency {translations}. But most interestingly, they bring
information that are not provided by other power-spectrum based
features, namely local direction of variation of power spectrum. 
As an illustration of this point, we will  compare the features obtained,
by MFCC and {the HOG-based approach on} two linear chirps, one with increasing frequency and the other one with a decreasing frequency, but both covering the same frequency range.  Our experimental results will show that bag of sole MFCC will
fail in fully capturing the discriminative information brought
by these signals at the contrary of the
features we propose.

\section{Data and Classifiers}
We provide in this section some details about the datasets
we have considered for evaluating the feature we propose.
Description of the classifier we used as well as
the experimental protocol are also given. 

\subsection{Toy dataset}
\begin{figure}[t]
\begin{center}
~\hfill
  \includegraphics[width=6cm]{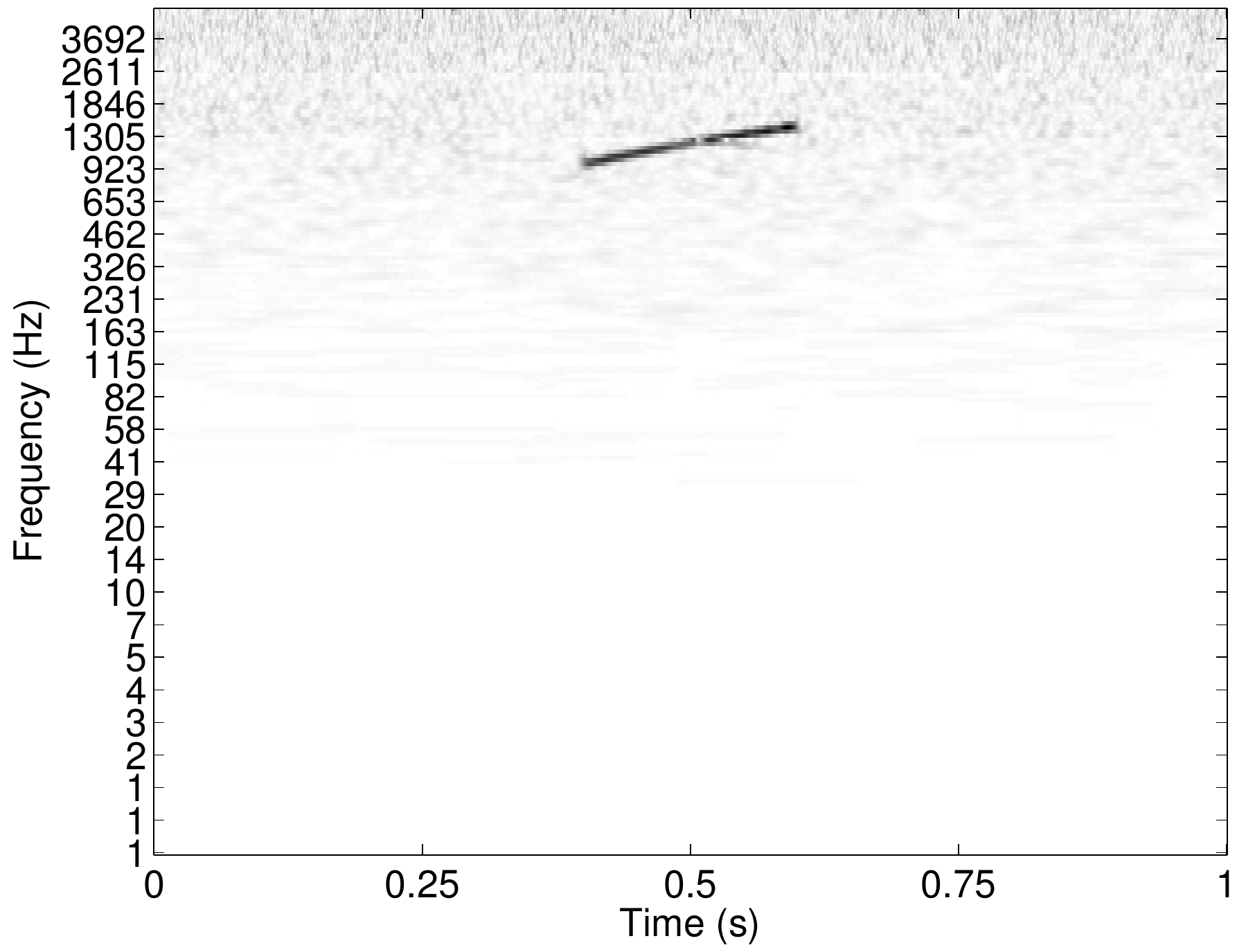}
\hfill~ \\ 
~\hfill  
\includegraphics[width=6cm]{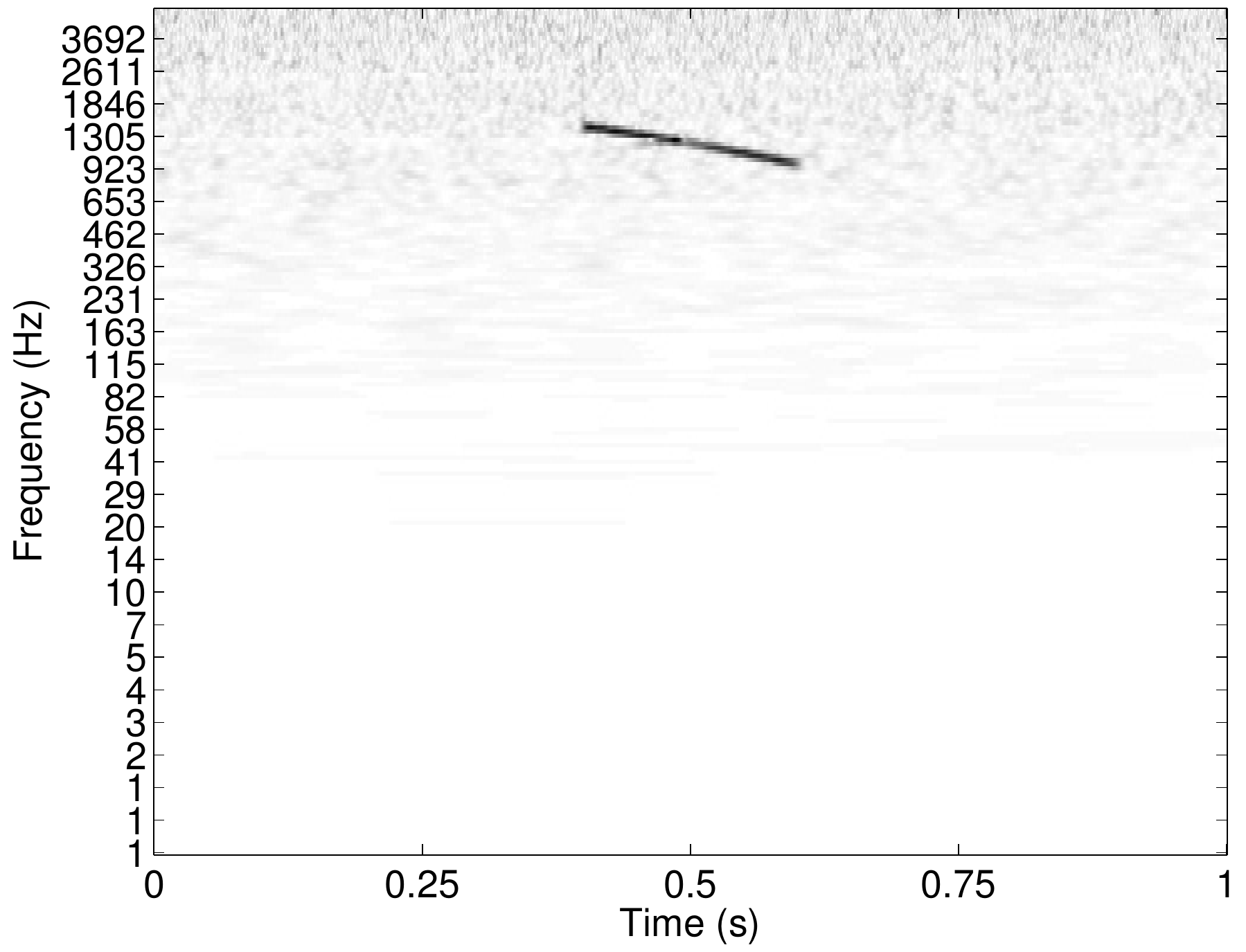}
\hfill~
\end{center}
  \caption{Examples of CQT representation of the two localized linear chirps used in the toy dataset. These images are obtained after a $512 \times 512$ resizing
and a $15 \times 15$  average filtering.}
  \label{fig:locchirp}
\end{figure}

For  evaluating  our features, we have created 
a toy problem which highlights the ability or the failure of studied features (including HOG and competitors)
in capturing power spectrum's direction of variation. As such,
we have created a binary classification problem where signals from
each class are composed of a localized linear chirp, respectively
of increasing and decreasing frequency, defined as
$$
s(t)=\Pi_{[t_1,t_2]}(t) \cos\big(2 \pi (a t + b) t \big) + n(t)
$$
\modifR{with $t \in [0,1]$} and  $n$ is a centered Gaussian noise of standard deviation $0.4$,
$a=1200$, $b=0$ for the class $y=+1$, $a=-1200$, $b=2400$ for the class $y=-1$
and $\Pi_{[t_1,t_2]}(t)$ a function which value is $1$ when $ t_1 \leq
t\leq t_2$ and $0$ otherwise. \modifR{We have set $t_1=0.4$ and
$t_2=0.6$.}
\modif{Figure \ref{fig:locchirp} depicts the CQ transform of representative samples of both classes. One can notice the localized spectral contents of the chirps.}

\subsection{D-case challenge dataset}

For {the} purpose of a challenge, a dataset providing environmental
sound recordings  has been recently
released by \cite{giannoulis13:_detec}. Each example in the dataset consists of  a $30$-second
audio scene, which has been captured at one of the $10$ following locations~:
\emph{bus, busy street, office, open air market, park, quiet street,
restaurant, supermarket, tube, tubestation}.
Recording has occurred at a rate of $44.1$ kHz and the number of 
examples available is $100$ with $10$ examples per class. 
Note that, the challenge's organisers have only made available
the development dataset\footnote{\url{http://c4dm.eecs.qmul.ac.uk/rdr/handle/123456789/29}}

\subsection{East Anglia {(EA)} dataset}
This dataset\footnote{ available at \url{ http://lemur.cmp.uea.ac.uk/Research/noise_db/}} has been collected in the early $2000$ by Ma et al.
\cite{ma03:_contex} at the East Anglia University. It provides  environmental
sounds coming from $10$ different
locations: \emph{bar, beach, bus, car, football match, laundrette,
lecture, office, railstation and street.} The length of each recording
is $4$ minutes and it has been recorded at a frequency of $22100$ Hz.
Similarly to the \emph{D-case} dataset, we have split the recording in
$30$-second audio scene {examples}. Hence, we have only $8$ examples
per class for this dataset.

\subsection{Litis Rouen dataset }

\begin{figure}[t]
\begin{center}
\begin{center}
\includegraphics[width=6cm]{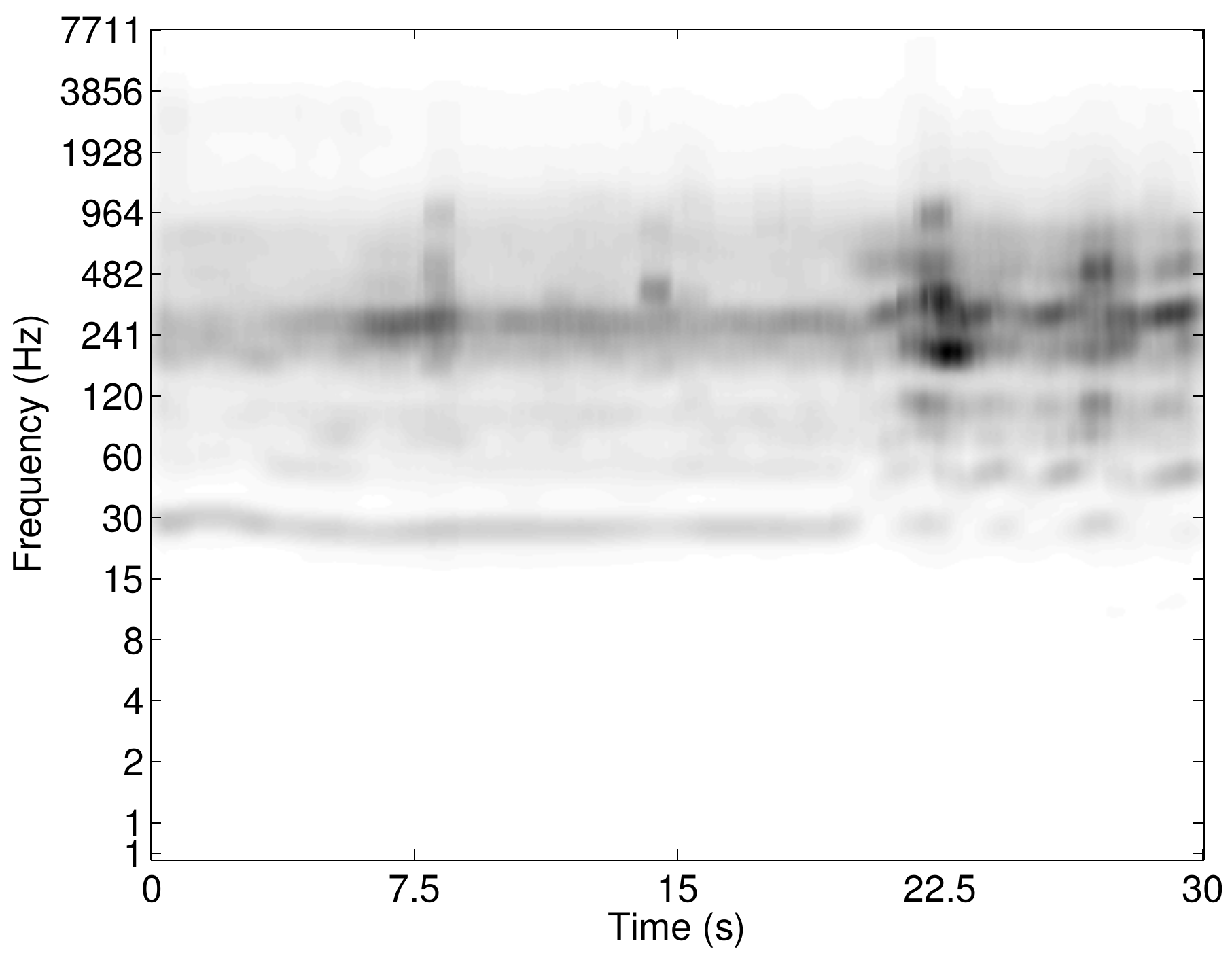}\\
(bus)
\end{center}   
\begin{center}
\includegraphics[width=6cm]{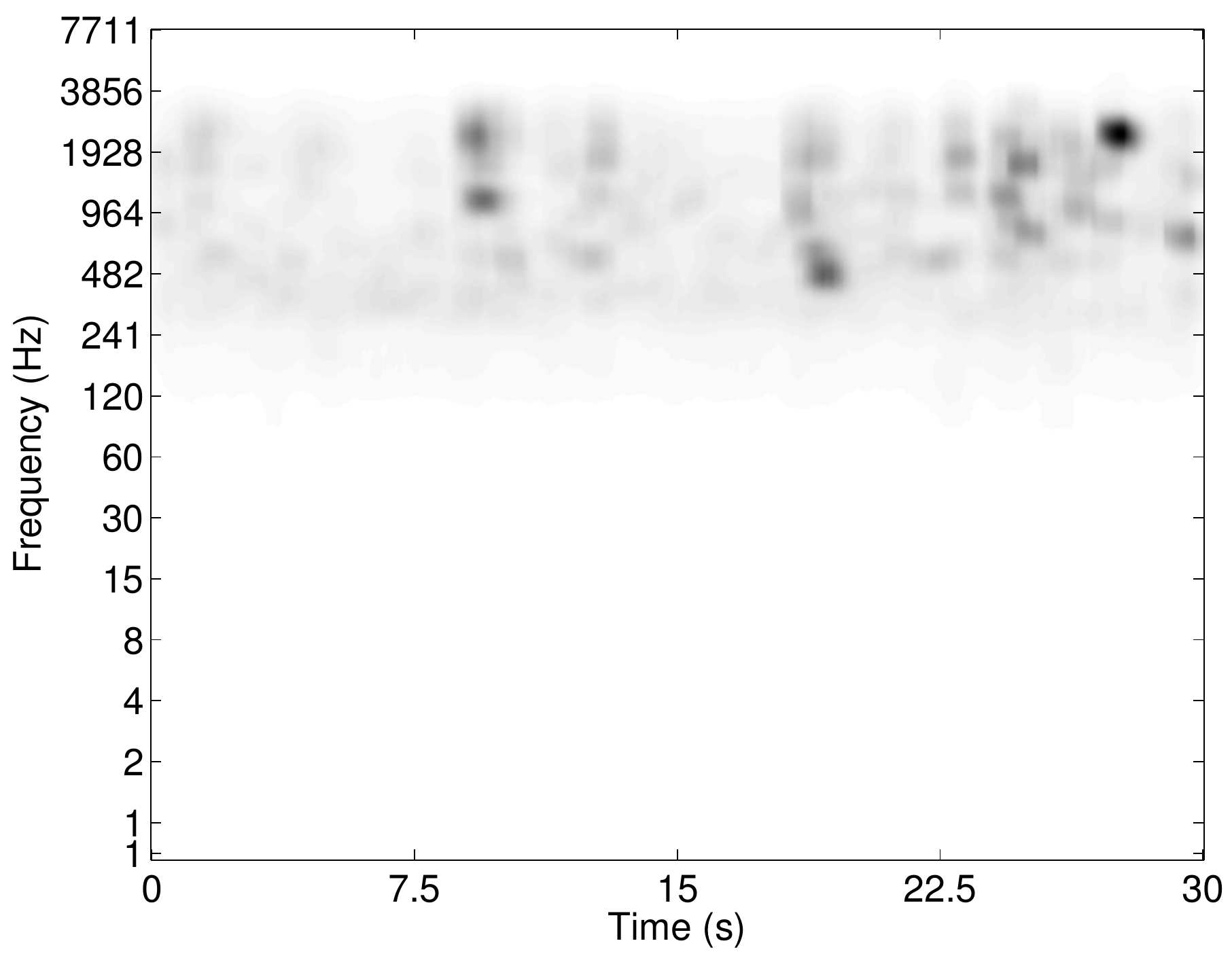}\\
(kid game hall)
\end{center}
\end{center}
  \caption{Examples of CQT of audio signals from $2$ different
scenes of the Rouen's dataset after image mean filtering
with an average kernel of size $15\times 15$.}
  \label{fig:cqtrouen}
\end{figure}

This dataset we make publicly available\footnote{available on the first author's website at : 
\url{https://sites.google.com/site/alainrakotomamonjy/home/audio-scene}
} goes beyond the above
ones in terms of volume and number of locations. Recordings
have been performed using a Galaxy S3 smartphone equipped 
with Android by means of the \emph{Hi-Q MP3 recorder} {application}.
While such an equipment may be considered as poor, we believe
that the resulting recordings would  be similar to
those obtained for {real applications} where cheap and {ubiquitous}
microphones are more likely to be used.    
The sampling frequency we have used is $44100$ Hz and the recording
is saved as a MP3 file with a bitrate of $64$ kbps.
When transformed into raw audio signals, they have been
downsampled to $22050$ Hz.
Overall, about $1500$ minutes of audio scene have been recorded.
They took place from December 2012 to June 2014.
\modif{For a given class, recordings  occured at several days of that period.
}
The dataset is composed of $19$ classes 
and audio scenes forming a given class have been
recorded at different locations.
 Note that in order to reduce temporal dependencies in our dataset,
 recordings usually last $1$ minute but in some locations, their
durations can reach up to $10$ minutes. 
Again in order to be consistent with the \emph{D-case} challenge,
each example is composed of a $30$-second audio scene. 
\modif{The 30-sec examples have
been obtained by splitting a given signal into 30-second segments without overlapping. 
}
A summary of the dataset is given
in Table \ref{tab:rouen}
and Figure \ref{fig:cqtrouen} presents some samples of
CQT for $2$ different audio scenes. The plots in this figure
show typical characteristics of an audio scene of the class. For instance,
in the \emph{bus}'s CQT, we can note the low-frequency line related
to the bus's acceleration and deceleration. In the \emph{kid game hall}
scene, we see some high-frequency structures induced by kid screams.

\begin{table}[t]
  \centering
  \caption{Summary of the Litis Rouen audio scene dataset}
  \label{tab:rouen}
  \begin{tabular}{l c}
\hline
   Classes &  \#examples\\ \hline\hline
plane  &   23\\\hline
busy street  &   143\\\hline
bus  &   192\\\hline
cafe  &   120\\\hline
car  &   243\\\hline
train station hall  &   269\\\hline
kid game hall  &   145\\\hline
market  &   276\\\hline
metro-paris  &   139\\\hline
metro-rouen  &   249\\\hline
biliard pool hall  &   155\\\hline
quiet street  &   90\\\hline
student hall  &   88\\\hline
restaurant  &   133\\\hline
pedestrian street  &   122\\\hline
shop  &   203\\\hline
train  &   164\\\hline
high-speed train  &   147\\\hline
tube station  &   125\\\hline\hline
  & 3026
\end{tabular}

\end{table}

\subsection{Competing features, classifier and protocols}

In order to evaluate how well the HOG-based feature we propose performs,
we have compared its performance to those of other features. As a sake of comparison, we
have considered the following ones:
\begin{itemize}
\item Bag of MFCC: these features are obtained
by computing the MFCC features on  windowed part
of the signals and then in concatenating them all \cite{aucouturier07}.  
The setting for the MFCC computations are typical. 
We have extracted MFCC features from each audio scene by means
of sliding windows of size $25$ ms with hops of $10$ ms. 
For each window, $13$ cepstra over $40$ bands have been computed. 
The toolbox we have employed is the \emph{rastamat} one with the 
dithering option on \cite{ellis05:_plp_rasta_mfcc_matlab}. The lower and upper frequencies 
of the spectral analysis are  respectively set to $1$ and
$10000$Hz. For the toy dataset, the upper frequency is set
to the Nyquist frequency. For obtaining
the final features, we average the obtained MFCC over batch of $40$ windows
with overlap of $20$ windows  and concatenated together
all the MFCC averages and standard deviations.
\modif{
\item Bag of MFCC-D-DD: these features are the same as above but in
addition to the MFCC averages and standard deviations over a batch of 
$40$ windows, we also concatenate to the feature vector, the average of first-order and second-order differences of the MFCC over these windows.
These features should provide a better description of the dynamic behaviour
of the signal.
\item \modifR{Frame/majority-based MFCC-D-DD~: For this approach, we use 
the typical solution which consists in computing
MFCCs and its first-order and second-order MFCC derivatives for
every  $10$ms-spaced frames of length $25$ms.   
 Each frame is described with a feature vector of size $39$ and is assigned
the label of the signal. Classifiers are trained on all frames and
at prediction time, a given $30$-second signal is labelled
according to the majority vote obtained over
the frames it is composed by.}
\item Texture-based TFR analysis: we have also implemented
the features extracted from time-frequency representations
as proposed in \cite{Yu:_Audio2009}.
These features exploit specific repeating spectral patterns in the TFR. 
Provided $M$ blocks (or filters) of different sizes in time-frequency
plane, feature generation proceeds by locally matching each of these blocks with
the CQT representation and retaining the highest local degree of
similarity of the block with the CQT image. This best local similarity
serves as feature, leading to an $M$-dimensional feature vector. In agreement
with \cite{Yu:_Audio2009}, the used local filters are randomly sampled over the CQT representations of the training set. For a strongly textured TFR, the retrieved filters are likely to be representative of the repeated patterns. However, 
for less-structured audio scene, the ability of this approach
to retrieve discriminative features may be spoiled.
In addition, these features
calculation is computationally demanding as it requires 2D
convolutions of the full TFR for the matching. In our experiments, we have
considered $M=20$ blocks of different sizes.
}
\item Recurrence plot analysis: these features are those introduced 
by \cite{roma13:_recur}, and they have achieved the best performance
on the test set of the D-case audio scene challenge \cite{giannoulis13:_detec}.
These are the features that we consider as the state-of-the-art.
The idea is to extract from  MFCC features, other characteristics that 
provide informations about recurrence over time of some specific
MFCC patterns. Interestingly, the final features proposed by \cite{roma13:_recur}  are obtained through averaging over time of all time-localized MFCC and
recurrence plot features. Hence, their features are of very low-dimensional
and do not provide any time-related information. 
MFCC features have been computed as above. Then, for all the MFCCs obtained over batch of $40$ windows with  overlap of $20$,  $11$ RQA features have been computed. 
Afterwards, MFCC features and RQA features are all averaged over time 
and MFCC averages, standard deviations and RQA averages are
concatenated to form a $37$-dimensional features. 

Note that we have considered an higher upper frequency of the spectral
analysis instead of the $900$Hz used by \cite{roma13:_recur}. Indeed,
we believe that their choice was optimal for one dataset at  the expense of genericity.

\end{itemize}

For our HOG feature, we have set the following parameters. The CQT
transform is computed, by means of the
\cite{schoerkhuber10:_const_q}'s  toolbox, on the same frequency range as the
MFCC features and with $8$ bins per octave. All other parameters have
been kept as default as proposed in
\cite{schoerkhuber10:_const_q}. The time-frequency representation is then
transformed into a $512\times 512$ image. The cells for
the histogram computation are of size $8 \times 8$ and we have chosen $8$ orientations. Note that it is also possible to  consider the signed direction
of gradient (leading thus to a histogram of size $16$). As described above,
histograms are normalized according to some norms, $4$  normalization factors 
are computed by the \emph{vlfeat} toolbox \cite{vedaldi08vlfeat} and we have considered the possibility
of using them as complementary features.\\

In order to compare our HOG-based feature to its competitors, we fed them to the same classifier and evaluated the resulting  performance. 
The classifier we have considered is an SVM classifier with either a linear
kernel or a Gaussian one. All problems except the toy one are multiclass classification problems. Hence, we have used a one-against-one scheme for dealing with this situation. 

 For all the experiments, we have provided averaged results
where the averaging occurs over $20$ different splits of the dataset into a
training set and a test set. For all datasets except the toy ones,
$80\%$ of the examples have been used for training. For the toy problem,
we considered only $40$ training examples among the $200$ available.
 Note that all features have been normalized so as to have zero
mean and unit variance on the training set. The test set has also been
normalized accordingly.
All the parameters of the SVM are tuned according to a validation
scheme. The $C$ parameter is selected among $10$ values
logarithmically scaled between $0.001$ and $100$ while the parameter
$\sigma$ of the Gaussian kernel $e^{-\frac{\|x -
    x^\prime\|^2}{2\sigma^2}}$ is chosen among $[1, 5, 10, 20, 50,
100]$.  Model selection is performed by resampling $5$ times, the training set
into a learning and validation set of equal size. The best hyperparameters are
considered as those maximizing averaged performances on the validation set.

As an evaluation criterion, we have considered the mean average precision, defined as
$$
MAP = \frac{1}{{N}}\sum_{i=1}^{N} \frac{TP(i)}{\#{N}(i)}
$$ 
where $TP(i)$ and $\#N(i)$ are respectively the number of examples of class $i$ correctly classified
and  the total number of examples classified in class $i$.

\section{Experimental results}
We have run several experiments aiming at showing the benefits
of our HOG-based feature compared to the state-of-the-art,
 as well as at analyzing the influence of the different parameters
of the HOG feature {extraction} pipeline.

\subsection{Comparison with classical features}

\begin{table*}[t]
  \centering
  \caption{Comparing performances of different features on the different datasets. Bold results  depict best performances
for each dataset as well as results   that are not statistically significantly different according to
a Wilcoxon signrank test with a p-value = $0.005$. Texture denotes the
features obtained from the work of Yu et al. \cite{Yu:_Audio2009}
while {MFCC-D-DD} is related to the MFCC and derivatives features.
{MFCC}, {MFCC-RQA}, {MFCC-900} and {MFCC-RQA-900} respectively {denote} the MFCC features, the MFCC and RQA features {at cut-off frequency of 10 KHz}, the MFCC and
the MFCC and RQA features with upper frequency set at $900$ Hz. The Hog (full)
and (marginalized) are related to the HOG features which are {respectively} obtained by concatenating the histograms from all cells and by concatenating
the two marginalized HOG features.
}
\begin{tabular}[h]{lll||cccc}
\multicolumn{3}{c}{}& \multicolumn{4}{c}{Datasets} \\\hline 
features & dim&  classifier & Toy & EA & D-Case & Rouen \\\hline \hline

Texture  & 20& linear   & \textbf{1.00 $\pm$ 0.00}   & 0.57 $\pm$ 0.13   & 0.23 $\pm$ 0.10 & -  \\\hline
Texture  & 20& gaussian   & \textbf{1.00 $\pm$ 0.01}   & 0.49 $\pm$ 0.13   & 0.19 $\pm$ 0.09 & -  \\\hline\hline
mfcc-d-dd  & 7800& linear   & \textbf{1.00 $\pm$ 0.00}   & \textbf{0.98 $\pm$ 0.04}   & 0.51 $\pm$ 0.13   & 0.66 $\pm$ 0.02   \\\hline
mfcc-d-dd  & 7800& gaussian   & \textbf{1.00 $\pm$ 0.01}   & 0.97 $\pm$ 0.05   & 0.49 $\pm$ 0.15   & 0.69 $\pm$ 0.02   \\\hline
\ar{mfcc-d-dd}  & \ar{39}& \ar{frame/majority}   & \ar{0.47 $\pm$ 0.04}   & \ar{0.95 $\pm$ 0.05}   & \ar{0.32 $\pm$ 0.12}   & \ar{0.36 $\pm$ 0.04}   \\\hline\hline
mfcc  & 3900& linear  & 0.78 $\pm$ 0.04   & \textbf{1.00 $\pm$ 0.01}   & 0.53 $\pm$ 0.12   & 0.67 $\pm$ 0.01   \\\hline
mfcc  & 3900& gaussian   & 0.77 $\pm$ 0.03   & \textbf{0.99 $\pm$ 0.03}   & 0.52 $\pm$ 0.10   & 0.76 $\pm$ 0.02   \\\hline
mfcc-900  & 3900& linear   & 0.50 $\pm$ 0.04   & 0.91 $\pm$ 0.07   & 0.53 $\pm$ 0.11   & 0.60 $\pm$ 0.02   \\\hline
mfcc-900  & 3900& gaussian   & 0.50 $\pm$ 0.04   & 0.86 $\pm$ 0.09   & 0.56 $\pm$ 0.12   & 0.66 $\pm$ 0.02   \\\hline\hline
mfcc+RQA  & 37& linear   & 0.51 $\pm$ 0.02   & 0.95 $\pm$ 0.08   & 0.54 $\pm$ 0.09   & 0.78 $\pm$ 0.01   \\\hline
mfcc+RQA  & 37& gaussian   & 0.51 $\pm$ 0.04   & 0.96 $\pm$ 0.08   & 0.52 $\pm$ 0.10   & \textbf{0.86 $\pm$ 0.02}   \\\hline
mfcc+RQA-900  & 37& linear   & 0.50 $\pm$ 0.04   & 0.93 $\pm$ 0.06   & \textbf{0.68 $\pm$ 0.10}   & 0.72 $\pm$ 0.02   \\\hline
mfcc+RQA-900  & 37& gaussian   & 0.49 $\pm$ 0.04   & 0.93 $\pm$ 0.07   & \textbf{0.65 $\pm$ 0.11}   & 0.80 $\pm$ 0.01   \\\hline\hline
Hog-full  & 65536& linear  & 0.97 $\pm$ 0.02   & \textbf{0.99 $\pm$ 0.02}   & 0.64 $\pm$ 0.09   & 0.84 $\pm$ 0.01   \\\hline
Hog-marginalized  & 2048& linear  & 0.94 $\pm$ 0.02   & \textbf{0.97 $\pm$ 0.06}   & \textbf{0.75 $\pm$ 0.10}   & 0.86 $\pm$ 0.01   \\\hline
Hog-marginalized  & 2048& gaussian   & 0.91 $\pm$ 0.04   & 0.95 $\pm$ 0.06   & \textbf{0.71 $\pm$ 0.12}   & \textbf{0.87 $\pm$ 0.01}   \\\hline
\end{tabular}

  \label{tab:compare}
\end{table*}

Our first result compares HOG features to some classical ones, the bag of MFCC,
the bag of MFCC-D-DD as described above as well as the features based on MFCC, Texture, Recurrence
quantitative analysis and the MFCC-D-DD frame-based classifier. We have considered two sets of signed HOG features. The first set, denoted as \emph{HOG-full},  concatenates all histograms obtained from all the cells, 
resulting  in a feature of very high-dimensionality. The 
second set of HOG features is obtained by averaging all the histograms
over the time and over the frequency and by concatenating the two
averaged histograms, denoted as \emph{HOG-marginalized}. Note that we have also used MFCC and MFCC-RQA
features obtained with an upper frequency of $900$ Hz as
in the paper of  \cite{roma13:_recur}.
Results obtained according to the above-described protocol and
the feature extraction parameters are depicted in 
Table \ref{tab:compare}. A Wilcoxon signed-rank test with
a p-value of $0.005$ have been computed between the 
best competitor and the {other} approaches.

\modif{For the toy dataset, the best performance is obtained
by the Texture and MFCC-D-DD based features with a perfect
classification. These features seem to be able to capture the
discriminative part of the signal either by capturing the MFCC variations 
or by finding the good  matching patch. We also explain
this good performance by the fact that the two classes
have high-energy well-localized time-frequency structures.}
{Except} for these features, our HOG-based feature performs
significantly better than its competitor RQA. MFCC is still able 
to discriminate  the two chirps but less powerfully than
{the} HOG feature, which natively {captures} the power spectrum variation.
Interestingly, the best performing HOG feature is the one 
which considers all the histograms, despite the very-high
dimensionality  and the low number
of training examples ($40$). A
rationale for this, is that the discriminative parts of the
signal are  very well-localized, thus
the   HOG features obtained from the cells covering this region 
are strongly discriminative. Of course, removing other spurious
HOG may have further increased performance.
\modifR{We can also remark that the use of the frame/majority-based classifier with 
MFCC-D-DD leads to poor results. This can be easily explained by
the fact that there are few discriminative frames in the audio scene
and this induces errors in the majority vote scheme. We thus believe
that such an approach is more appropriate to audio classification
problems where
discriminative features are persistent over time like in music
genre classification}. We will show in the sequel that when using appropriate 
cell size, our HOG feature will also obtain perfect classification.

The \emph{East Anglia}'s dataset seems to be fairly easy and all
features except the \ar{Texture-based  and  frame/majority approaches}
  perform well with a slight advantage to MFCC.

\modif{For the \emph{D-case} challenge,
the Texture, MFCC, MFCC-D-DD and MFCC-RQA features
perform poorly, with performances around $50\%$ and even around $20\%$ for the Texture}. However, with a more adapted upper-frequency of
the spectral analysis, performances of the MFCC-RQA reach $68\%$ of 
mean average precision. Marginalized {HOG} features perform
significantly better than competitors with a gain in performance
of about $19\%$ when the  range of frequency 1-10000 Hz is considered. {This gain drops to 7\% but is still consequent when the range of frequency 1-900 Hz is used for MFCC-RQA.}
 However, we can note that using
the full HOG representation induces
a slight loss of performances and that it seems
 valuable to consider some HOG pooling. 
We also want to highlight that the performance we report
for MFCC-RQA-900 is slightly lower than those given in 
\cite{roma13:_recur} and this is due to the fact
our results are  averages over $20$ splits
instead of a $5$-fold cross-validation precision.

For the \emph{Rouen}'s dataset we have introduced, the marginalized HOG features
 perform on par with MFCC-RQA  and
 significantly better than all other competitors.
Using a cut-off frequency of $900$ Hz induces
a larger loss of performance for {these} MFCC-RQA features.
Also note that we cannot provide results using
the Texture-based features as after one week,
the feature computation was still running proving that approach is intractable even
for medium-scale datasets.
\modifR{Again, MFCC-D-DD features with the frame/majority-based
classifier  fail to capture
the discriminative information of the audio scene}.

More interestingly, we highlight that the  marginalized HOG feature
is robust across the different datasets, especially when used
in conjunction with a linear kernel,  even though its {extraction} parameters
have not been tuned. This is a very promising result concerning
the generalization capability of these features. We
will see in the sequel that by properly tuning the HOG  parameters
we are able to statistically significantly perform better than the MFCC-RQA feature on the Rouen's dataset.

\subsection{Analyzing HOG feature parameters}

\begin{table*}[t]
  \centering
  \caption{Analyzing the effects of HOG  feature parameters. Two different
parameters have been evaluated: the sign of gradient orientations in the histogram
computations (signed, unsigned and both) { and  the}
inclusion (with or without) of the normalization factors. Bold results depict best performances
for each dataset as well as results   that are not statistically significantly different according to
a Wilcoxon signrank test with a p-value = $0.005$.}
  \label{tab:param}
  \begin{tabular}[t]{ccc||cccc}
\multicolumn{3}{c}{}& \multicolumn{4}{c}{Datasets} \\\hline 
sign & factors  &  dim & Toy & EA & {D-case} & Rouen \\\hline \hline

signed  & w/o & 2048   & 0.94 $\pm$ 0.02   & \textbf{0.97 $\pm$ 0.06}   & \textbf{0.75 $\pm$ 0.10}   & 0.86 $\pm$ 0.01   \\\hline
signed  & with & 2560   & 0.91 $\pm$ 0.02   & \textbf{0.97 $\pm$ 0.08}   & \textbf{0.73 $\pm$ 0.11}   & \textbf{0.86 $\pm$ 0.02}   \\\hline
unsigned  & w/o & 1024   & \textbf{0.97 $\pm$ 0.01}   & \textbf{0.97 $\pm$ 0.07}   & \textbf{0.70 $\pm$ 0.11}   & 0.83 $\pm$ 0.02   \\\hline
unsigned  & with & 1536   & 0.93 $\pm$ 0.02   & \textbf{0.97 $\pm$ 0.08}   & 0.67 $\pm$ 0.11   & 0.84 $\pm$ 0.02   \\\hline
both  & w/o & 3072   & \textbf{0.96 $\pm$ 0.01}   & \textbf{0.98 $\pm$ 0.04}   & \textbf{0.75 $\pm$ 0.12}   & \textbf{0.86 $\pm$ 0.01}   \\\hline
both  & with & 3584   & 0.95 $\pm$ 0.02   & \textbf{0.98 $\pm$ 0.06}   & \textbf{0.75 $\pm$ 0.11}   & \textbf{0.86 $\pm$ 0.01}   \\\hline

  \end{tabular}

\end{table*}

\begin{table*}[t]
  \centering
  \caption{Analyzing the effects of HOG  feature parameters and the image filtering parameter. Three different
parameters have been evaluated: the size of the cell for normalization, the number of orientations
in the HOG computation and the average image filtering size. Bold results depict best performances
for each dataset as well as results   that are not statistically significantly different according to
a Wilcoxon signrank test with a p-value = $0.005$.}
  \label{tab:cellfilter}
  \begin{tabular}[t]{cccc||cccc}
\multicolumn{4}{c}{}& \multicolumn{4}{c}{Datasets} \\\hline 
filter size & Nb. Orient  &  cell size & dim  & Toy & EA & {D-case} & Rouen \\\hline \hline
15 & 8 & 2 & 12288  & \textbf{1.00 $\pm$ 0.00}   & \textbf{0.97 $\pm$ 0.05}   & 0.66 $\pm$ 0.13   & 0.84 $\pm$ 0.01   \\\hline
15 & 8 & 4 & 6144  & 0.98 $\pm$ 0.01   & \textbf{0.98 $\pm$ 0.05}   & \textbf{0.73 $\pm$ 0.12}   & 0.85 $\pm$ 0.01   \\\hline
15 & 8 & 8 & 3072  & 0.96 $\pm$ 0.01   & \textbf{0.98 $\pm$ 0.04}   & \textbf{0.75 $\pm$ 0.12}   & 0.86 $\pm$ 0.01   \\\hline
15 & 8 & 16 & 1536  & 0.98 $\pm$ 0.01   & \textbf{0.96 $\pm$ 0.07}    & \textbf{0.73 $\pm$ 0.15}   & 0.86 $\pm$ 0.02   \\\hline
15 & 8 & 32 & 768  & \textbf{1.00 $\pm$ 0.00}   & \textbf{0.98 $\pm$ 0.04}   & 0.68 $\pm$ 0.12   & 0.83 $\pm$ 0.02   \\\hline
1 & 8 & 8 & 3072  & 0.97 $\pm$ 0.01   & \textbf{0.97 $\pm$ 0.05}   & \textbf{0.65 $\pm$ 0.12}   & \textbf{{0.89 $\pm$ 0.01}}\\\hline
5 & 8 & 8 & 3072  & 0.96 $\pm$ 0.01   & \textbf{0.99 $\pm$ 0.01}   & \textbf{0.69 $\pm$ 0.14}   & \textbf{0.88 $\pm$ 0.01}   \\\hline
15 & 8 & 8 & 3072  & 0.96 $\pm$ 0.01   & \textbf{0.98 $\pm$ 0.04}   & \textbf{0.75 $\pm$ 0.12}   & 0.86 $\pm$ 0.01   \\\hline
30 & 8 & 8 & 3072  & 0.99 $\pm$ 0.01   & \textbf{0.96 $\pm$ 0.07}   & \textbf{0.69 $\pm$ 0.12}   & 0.82 $\pm$ 0.02   \\\hline
15 & 4 & 8 & 1536  & 0.79 $\pm$ 0.03   & \textbf{0.98 $\pm$ 0.05}   & \textbf{0.74 $\pm$ 0.12}   & 0.86 $\pm$ 0.01   \\\hline
15 & 8 & 8 & 3072  & 0.96 $\pm$ 0.01   & \textbf{0.98 $\pm$ 0.04}   & \textbf{0.75 $\pm$ 0.12}   & 0.86 $\pm$ 0.01   \\\hline
15 & 16 & 8 & 6144  & 0.98 $\pm$ 0.01   & \textbf{0.97 $\pm$ 0.07}   & \textbf{0.76 $\pm$ 0.10}   & 0.86 $\pm$ 0.02   \\\hline
  \end{tabular}

\end{table*}

In the first part of this experiment, we have investigated the influence of two parameters of the HOG features on the global classification performance.  We have used the  marginalized HOG features, as in the previous experiment, in {conjunction} with a linear kernel. The HOG features are {composed of} either signed, unsigned or both histograms eventually completed with the $4$ normalization factors. Hence for each cell, the size of the feature ranges from $8$ to
$28= 8 \times 3 + 4$. Since we have $64$ rows and columns of $8\times8$ cells
in the image, this results in feature vector of size ranging from $1024$ to $3584$.

The results we obtain for the different datasets are presented in Table \ref{tab:param}. We can note that the parameters we are evaluating clearly
influence performances. Depending on the datasets, the variation of
{performances} is in between $2\%$ (for \emph{East Anglia}) to $6\%$
(for \emph{D-case} and the toy dataset). The most consistent feature seems to be the ones
for which  histograms are computed with both signed and unsigned
gradient orientations and  the normalization factors are not
included.

\modif{In the second part of the experiment, we have analyzed the effect of the average kernel size of the mean
filtering, the number of orientations in the HOG as well as the size of the cell.{The results are reported in Table \ref{tab:cellfilter}}.
We note that these parameters have different influences given the dataset. However, it seems
that the choices of a cell size of $8$ and $8$ orientations for the HOG computation are
a good default choice. The parameter with most influence seems to be
the filtering kernel size. Depending on the datasets, the best kernel size varies
from $1$, which corresponds to no filtering to $15$. According to
this result, we thus suggest practictioner to 
 tune this parameter value according to its own dataset.
 }

\subsection{On the effect of pooling}

\begin{table*}[t]
  \centering
  \caption{Analyzing the effects of pooling. The number under
the \emph{Freq} and \emph{Time} labels depicts the number of histograms
on the frequency and time axes after pooling. For instance, the first row
presents the result of pooling where all histograms {have}
been averaged over the frequency axes. This pooling corresponds
to the red pooling in Figure \ref{fig:pooling}. Bold results depict best performances
for each dataset as well as results   that are not statistically significantly different according to
a Wilcoxon signrank test with a p-value = $0.005$.
   }
  \label{tab:block}
  \begin{tabular}[t]{ccc||cccc}
\multicolumn{3}{c}{}& \multicolumn{4}{c}{Datasets} \\\hline 
Freq & Time &  dim & Toy & EA & D-case & Rouen \\\hline \hline
1  & 64& 1536   & \textbf{0.99 $\pm$ 0.01}   & 0.69 $\pm$ 0.10   & 0.43 $\pm$ 0.11   & 0.41 $\pm$ 0.02   \\\hline
2  & 32& 1536   & 0.98 $\pm$ 0.02   & 0.87 $\pm$ 0.07   & 0.54 $\pm$ 0.15   & 0.56 $\pm$ 0.02   \\\hline
4  & 16& 1536   & 0.96 $\pm$ 0.02   & 0.92 $\pm$ 0.08   & 0.56 $\pm$ 0.13   & 0.68 $\pm$ 0.02   \\\hline
8  & 8& 1536   & 0.93 $\pm$ 0.02   & 0.97 $\pm$ 0.05   & 0.64 $\pm$ 0.14   & 0.78 $\pm$ 0.01   \\\hline
16  & 4& 1536   & \textbf{0.99 $\pm$ 0.01}   & \textbf{0.97 $\pm$ 0.06}   & \textbf{0.70 $\pm$ 0.12}   & 0.85 $\pm$ 0.01   \\\hline
32  & 2& 1536   & 0.98 $\pm$ 0.01   & \textbf{0.99 $\pm$ 0.04}   & \textbf{0.75 $\pm$ 0.12}   & 0.88 $\pm$ 0.01   \\\hline
64  & 1& 1536   & \textbf{0.99 $\pm$ 0.01}   & \textbf{1.00 $\pm$ 0.01}   & \textbf{0.73 $\pm$ 0.10}   & \textbf{0.92 $\pm$ 0.01}   \\\hline

\end{tabular}
\end{table*}

We have analyzed the effects of pooling on the
performances of the HOG features. Indeed, it is well known from
the computer vision literature that pooling plays an important
role when it comes to pattern recognition \cite{boureau10} and
we believe that a proper choice of pooling can also improve
performances in our audio scene classification problem.
In the experiment, we  varied the size of the average pooling in the time and frequency axis. 
Here, the HOG features is obtained using both signed and
unsigned histograms and without the normalization factors.
\modif{The mean filtering size, the number of cells and the number
of orientations have been set to the values that maximize
performances according to Table \ref{tab:cellfilter}.}
Again, linear kernel is used in the SVM. Note that the form of HOG pooling investigated
in this experiment does not necessarily perform better than those used in previous ones.

Results for different sizes of pooling are presented in Table
\ref{tab:block}. Note that the first and last rows correspond
respectively to the results obtained for the red and green pooling in
Figure \ref{fig:pooling}. Other rows are related to more general
pooling form as in the blue pooling in Figure \ref{fig:pooling}.  A
striking result can first be highlighted, regarding the importance of
carefully selecting the pooling form~: variation of performance
between the worst and the best pooling form is at least $30\%$ for all
real datasets.

Worst performance is achieved by pooling over frequency, which means
that we average all the obtained histograms over the frequency, losing
all informations about spectral contents. This finding is  rather intuitive as we believe that
the audio scenes can be mostly discriminated by their spectral contents
and the local variations of their spectral contents.

At the other end, best performances are obtained by pooling over time,
especially when we consider the real datasets. This result is also
interesting in the sense that best performances are achieved while no
time information are kept in the features as they are totally
translation-invariant. We make the hypothesis that this occurs because
most audio scenes  can be distinguished
according to some global analysis (enhanced by pooling over time) of
some recurrent patterns without the needs to look at some short-time
single events, although these events may carry discriminative
information.  This rationale  is also corroborated by the
fact that MFCC-RQA features that are averaged over time performs
reasonably well on the real datasets. While this approach works
pretty well for the audio scene classes we have considered, we
believe that features able to leverage on short-time
events will be needed for fine-grained audio-scene classification.

Regarding other pooling forms, we can note a clear trend of
improving performances as the pooling over frequency is decreased
and the one over time increases.

\subsection{More insights on the Rouen's dataset}
\begin{table*}[t]
  \centering
  \caption{Normalized sum  of all confusion matrices obtained over the 
$20$ training/test splits. The normalization occurs by columns so that the diagonal
terms represent the precision obtained for a given class. The obtained
average precision is $0.915$.
 The HOG features we used are the best performing ones
according to above experiments. Rows depict the real class of the audio scene
while Columns are related to the predicted one.
  }

\tiny
  \begin{tabular}[t]{l||ccccccccccccccccccc}
& \rot{plane}& \rot{busy street}& \rot{bus}& \rot{cafe}& \rot{car}& \rot{train station hall}& \rot{kid game hall}& \rot{market}& \rot{metro-paris}& \rot{metro-rouen}& \rot{billiard pool hall}& \rot{quiet street}& \rot{student hall}& \rot{restaurant}& \rot{pedestrian street}& \rot{shop}& \rot{train}& \rot{ high-speed train}& \rot{tubestation}\\\hline
plane& 97.0& 0.0& 0.0& 0.0& 0.0& 0.4& 0.0& 0.0& 0.0& 0.0& 0.0& 0.0& 0.0& 0.0& 0.0& 0.0& 0.0& 0.0& 0.0\\\hline
busy street& 0.0& 83.2& 0.0& 2.1& 0.0& 0.1& 1.3& 0.9& 1.0& 2.4& 0.0& 6.1& 0.0& 0.0& 0.2& 0.0& 0.0& 0.0& 0.0\\\hline
bus& 0.0& 0.0& 98.6& 0.0& 0.0& 0.0& 0.0& 0.0& 0.2& 0.3& 0.0& 0.0& 0.0& 0.0& 0.0& 0.1& 0.0& 0.0& 0.0\\\hline
cafe& 0.0& 1.0& 0.0& 82.9& 0.0& 0.2& 0.0& 0.6& 0.0& 0.0& 0.0& 1.9& 0.0& 0.2& 11.3& 0.6& 0.0& 0.0& 0.0\\\hline
car& 0.0& 0.0& 0.1& 0.0& 99.9& 0.0& 0.0& 0.0& 0.0& 0.0& 0.0& 0.0& 0.0& 0.0& 0.0& 0.0& 0.0& 0.0& 0.0\\\hline
train station hall& 0.0& 0.2& 0.0& 0.2& 0.0& 92.9& 1.2& 1.7& 0.2& 0.0& 0.0& 0.0& 0.0& 0.2& 2.2& 0.0& 0.0& 0.0& 0.0\\\hline
kid game hall& 0.0& 0.0& 0.0& 0.0& 0.0& 0.0& 96.7& 0.0& 0.0& 0.0& 0.0& 0.0& 0.0& 0.0& 0.0& 0.0& 0.0& 0.0& 0.0\\\hline
market& 0.0& 0.3& 0.0& 1.4& 0.0& 1.3& 0.0& 90.5& 0.0& 0.0& 0.0& 0.6& 0.0& 0.7& 4.6& 1.4& 0.0& 0.0& 0.2\\\hline
metro-paris& 0.0& 2.8& 0.5& 0.2& 0.0& 0.2& 0.0& 0.0& 86.7& 7.2& 0.0& 1.3& 0.0& 0.0& 0.2& 0.6& 0.0& 0.0& 0.0\\\hline
metro-rouen& 3.0& 2.2& 0.3& 0.6& 0.0& 0.0& 0.7& 0.0& 10.0& 88.3& 0.0& 0.3& 0.0& 0.0& 0.0& 0.0& 0.6& 0.0& 0.0\\\hline
billiard pool hall& 0.0& 0.0& 0.0& 0.0& 0.0& 0.0& 0.0& 0.1& 0.0& 0.0& 100.0& 0.0& 0.0& 0.0& 0.7& 0.2& 0.0& 0.0& 0.0\\\hline
quiet street& 0.0& 7.5& 0.0& 3.5& 0.0& 0.0& 0.0& 0.0& 0.6& 0.2& 0.0& 75.9& 0.3& 0.0& 3.9& 4.3& 0.0& 0.0& 0.4\\\hline
student hall& 0.0& 0.0& 0.0& 0.0& 0.0& 0.0& 0.0& 0.0& 0.0& 0.0& 0.0& 0.0& 98.5& 2.1& 1.7& 1.3& 0.0& 0.0& 0.0\\\hline
restaurant& 0.0& 0.0& 0.0& 0.2& 0.0& 0.2& 0.0& 0.8& 0.0& 0.0& 0.0& 0.0& 0.9& 92.4& 0.0& 0.6& 0.0& 0.0& 0.0\\\hline
pedestrian street& 0.0& 1.2& 0.0& 7.2& 0.0& 2.8& 0.0& 2.5& 1.0& 0.0& 0.0& 5.1& 0.0& 1.1& 70.9& 2.9& 0.0& 0.0& 0.2\\\hline
shop& 0.0& 0.2& 0.0& 1.2& 0.0& 0.6& 0.2& 2.5& 0.0& 0.3& 0.0& 5.1& 0.3& 3.2& 3.3& 86.4& 0.0& 0.0& 0.2\\\hline
train& 0.0& 0.0& 0.5& 0.0& 0.1& 0.6& 0.0& 0.0& 0.2& 1.1& 0.0& 0.0& 0.0& 0.0& 0.0& 0.0& 99.1& 0.0& 0.0\\\hline
 high-speed train& 0.0& 0.0& 0.0& 0.0& 0.0& 0.0& 0.0& 0.0& 0.2& 0.3& 0.0& 0.0& 0.0& 0.0& 0.0& 0.0& 0.3& 100.0& 0.0\\\hline
tubestation& 0.0& 1.5& 0.0& 0.4& 0.0& 0.8& 0.0& 0.4& 0.0& 0.0& 0.0& 3.5& 0.0& 0.2& 1.1& 1.6& 0.0& 0.0& 98.9\\\hline
  \end{tabular}

  \label{tab:confmatrirouen}
\end{table*}
As one of our main contribution in this paper is to introduce
a novel audio scene dataset, we discuss in the sequel our findings
regarding this dataset. 

In Table \ref{tab:block}, we have shown that  mean average precision, obtained as an average over
$20$ trials, is $0.9170$. Table \ref{tab:confmatrirouen} presents the normalized sum of all confusion matrices
obtained from these $20$ training/test splits. They have been obtained 
using the best performing HOG feature~: namely the one with signed
and unsigned orientations, without normalization factors {of the} histograms
and fully pooled over time, the HOG being obtained with cell size of $8$,
$8$ bins of the orientations and no average filtering. 
The average precision obtained from this matrix is $0.915$ and it is different to the mean average precision over the
$20$ runs as presented in Table \ref{tab:block}.

\begin{figure*}[t]
\begin{center}
~\hfill
  \includegraphics[width=3.5cm]{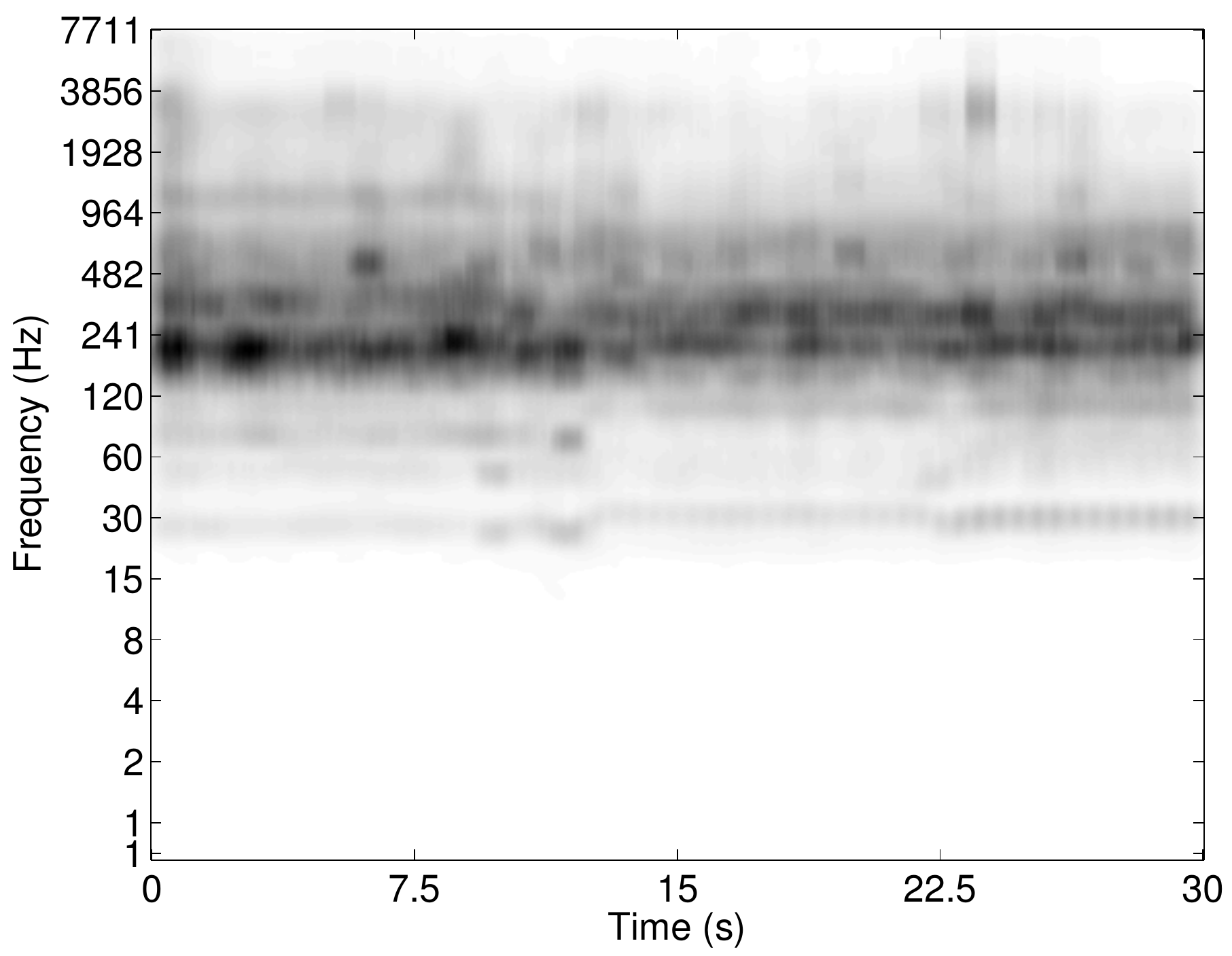}
~\hfill~ 
\includegraphics[width=3.5cm]{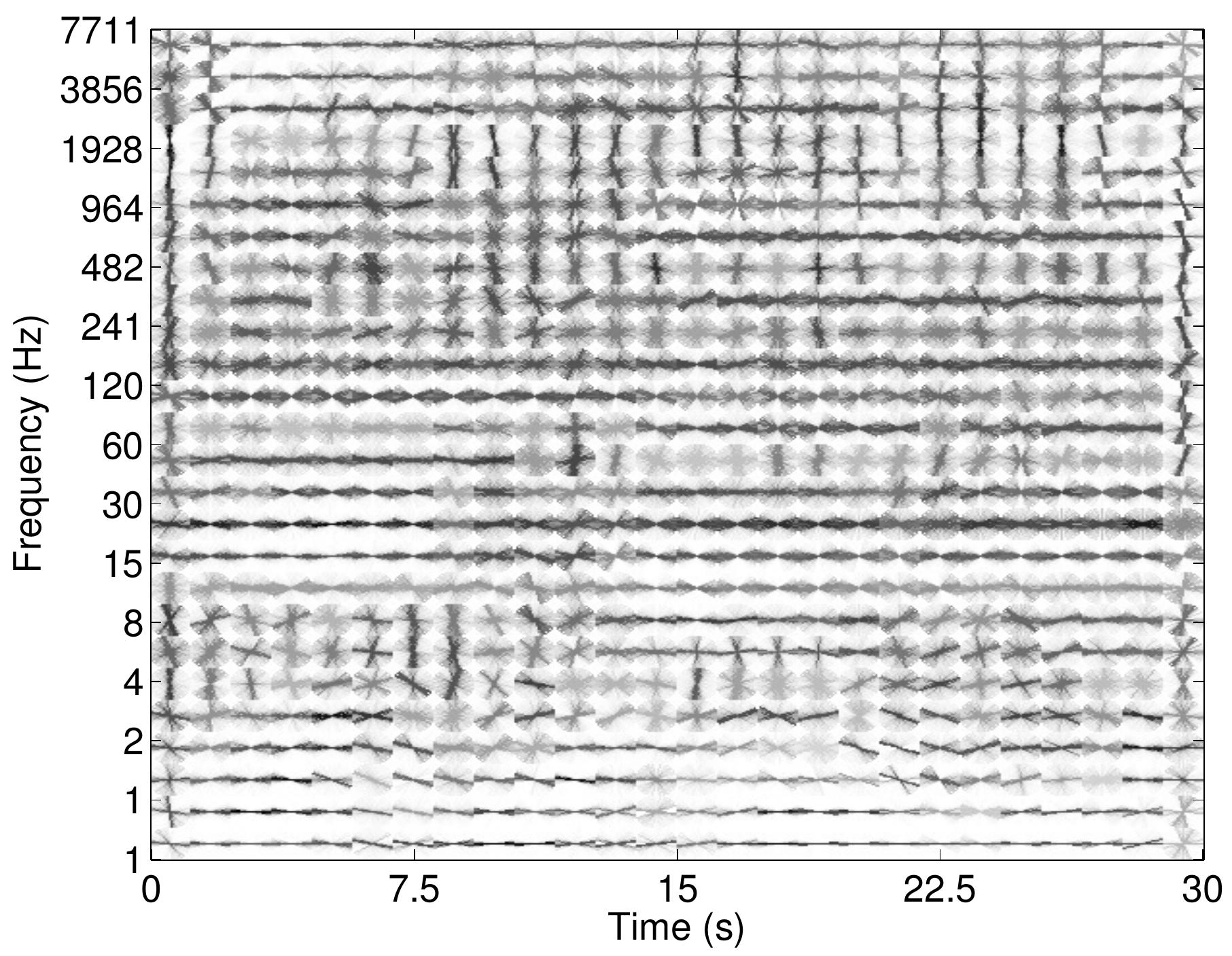}
\hfill~
\includegraphics[width=3.5cm]{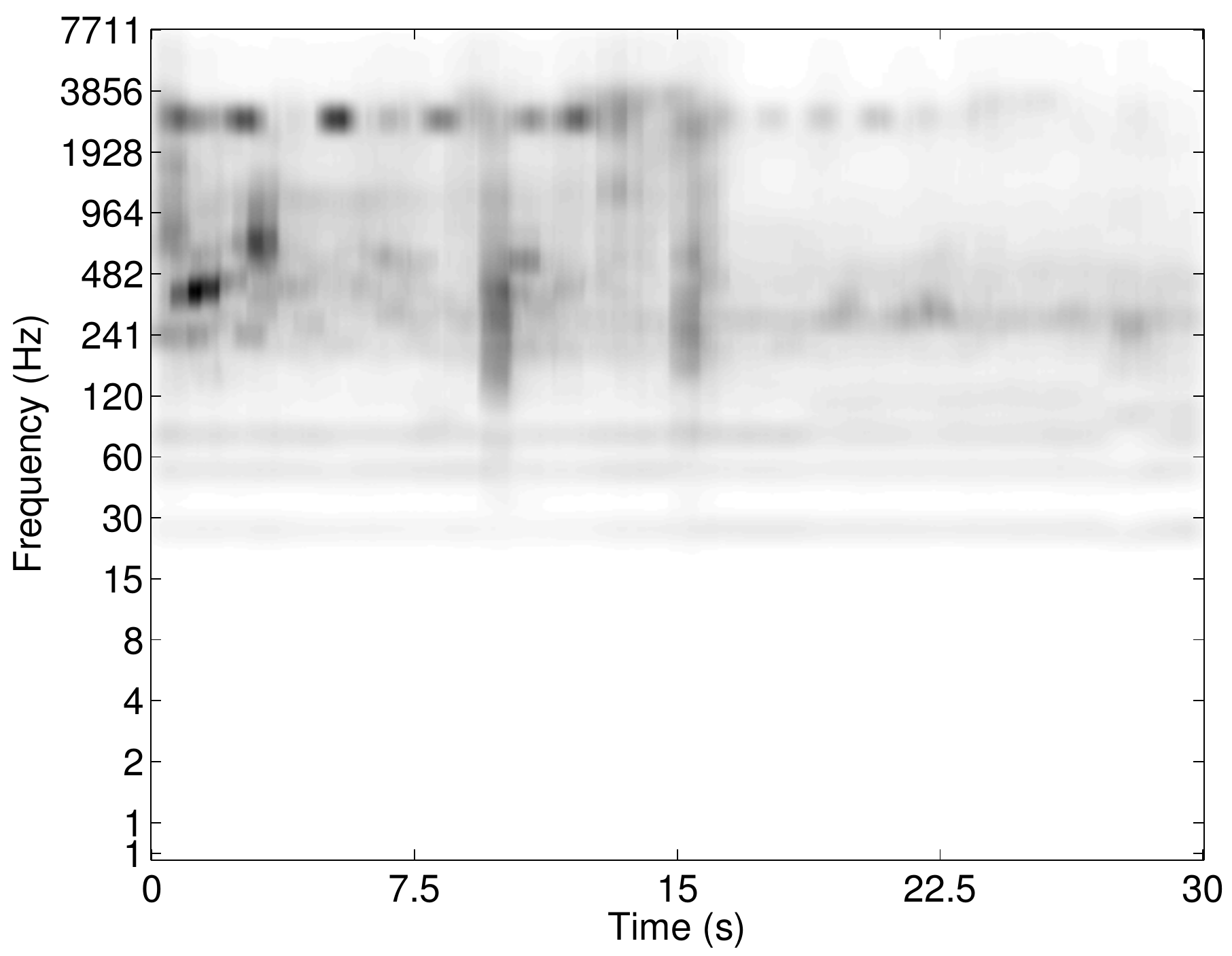}
~\hfill~ 
\includegraphics[width=3.5cm]{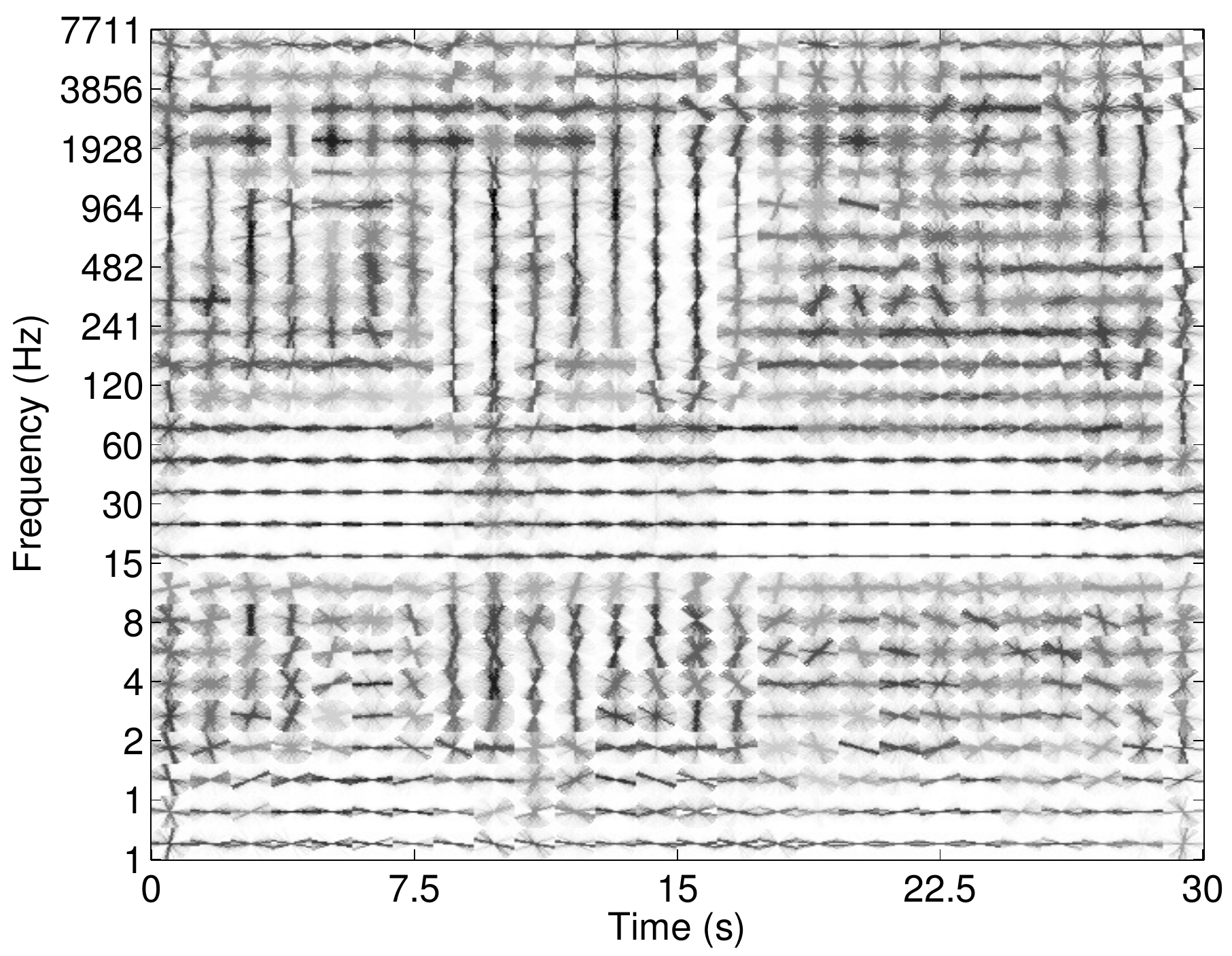}
\hfill~ \\
~\hfill (bus) \hfill~ \\
\vspace{0.3cm}
~\hfill
  \includegraphics[width=3.5cm]{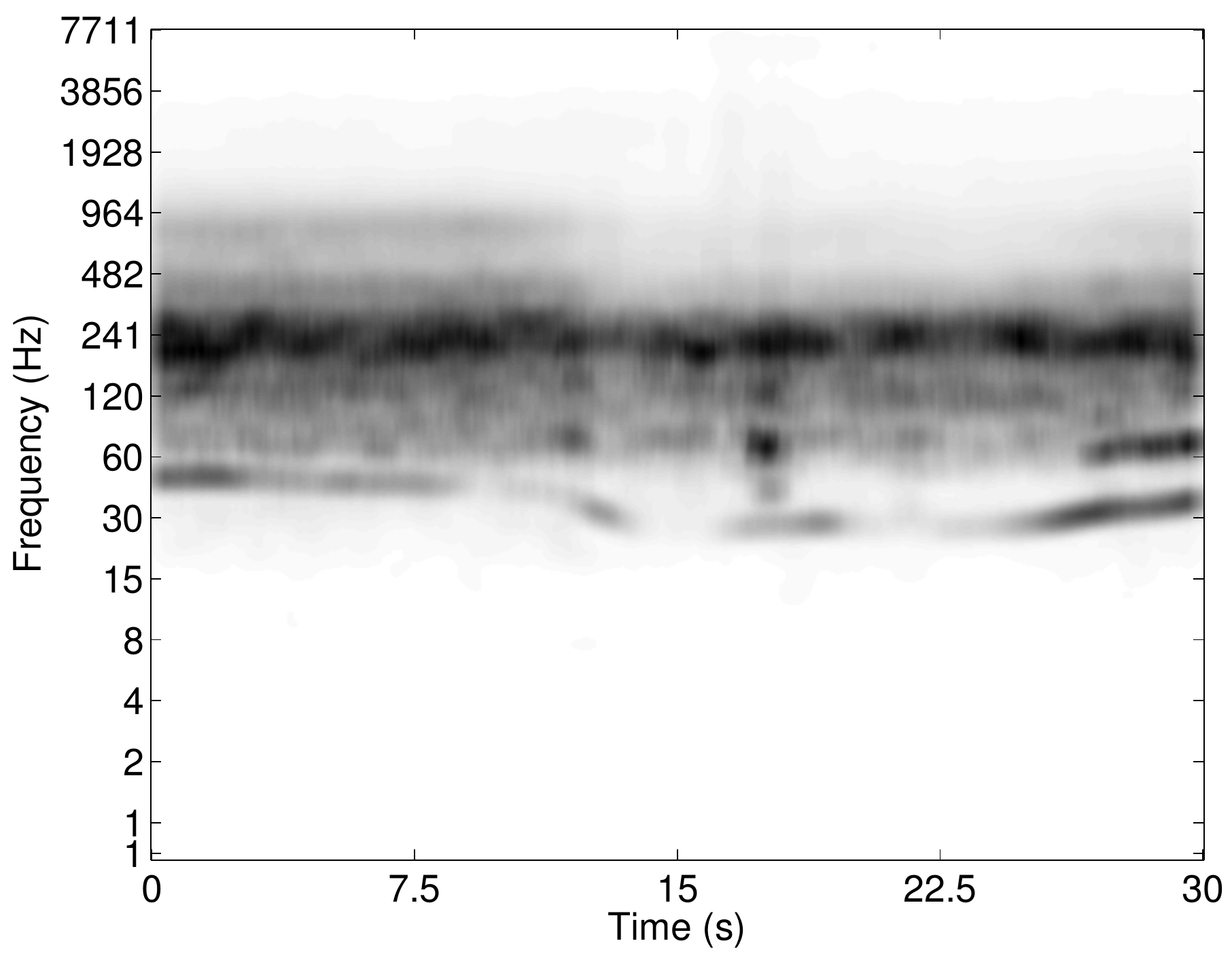}
~\hfill~ 
\includegraphics[width=3.5cm]{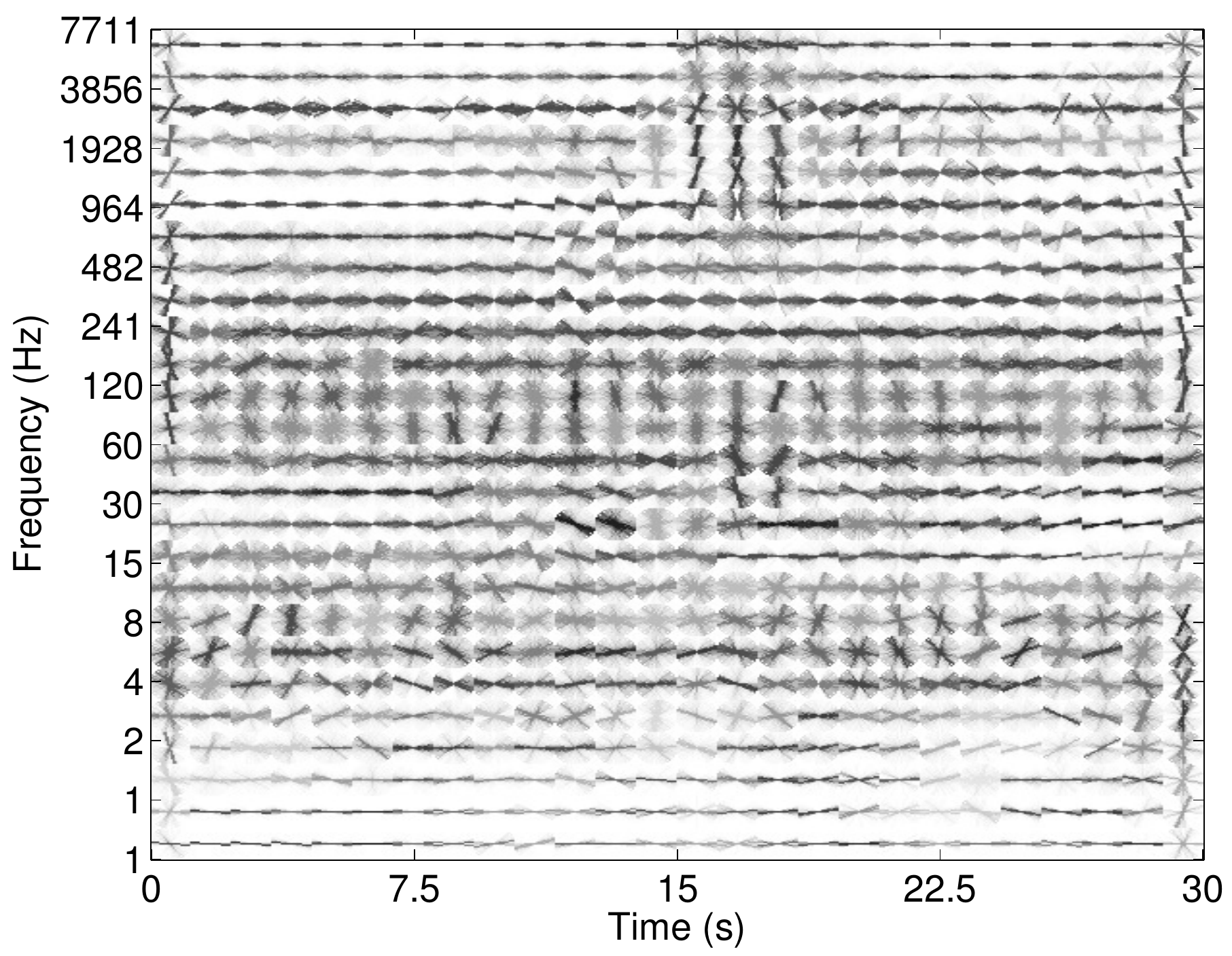}
\hfill~
\includegraphics[width=3.5cm]{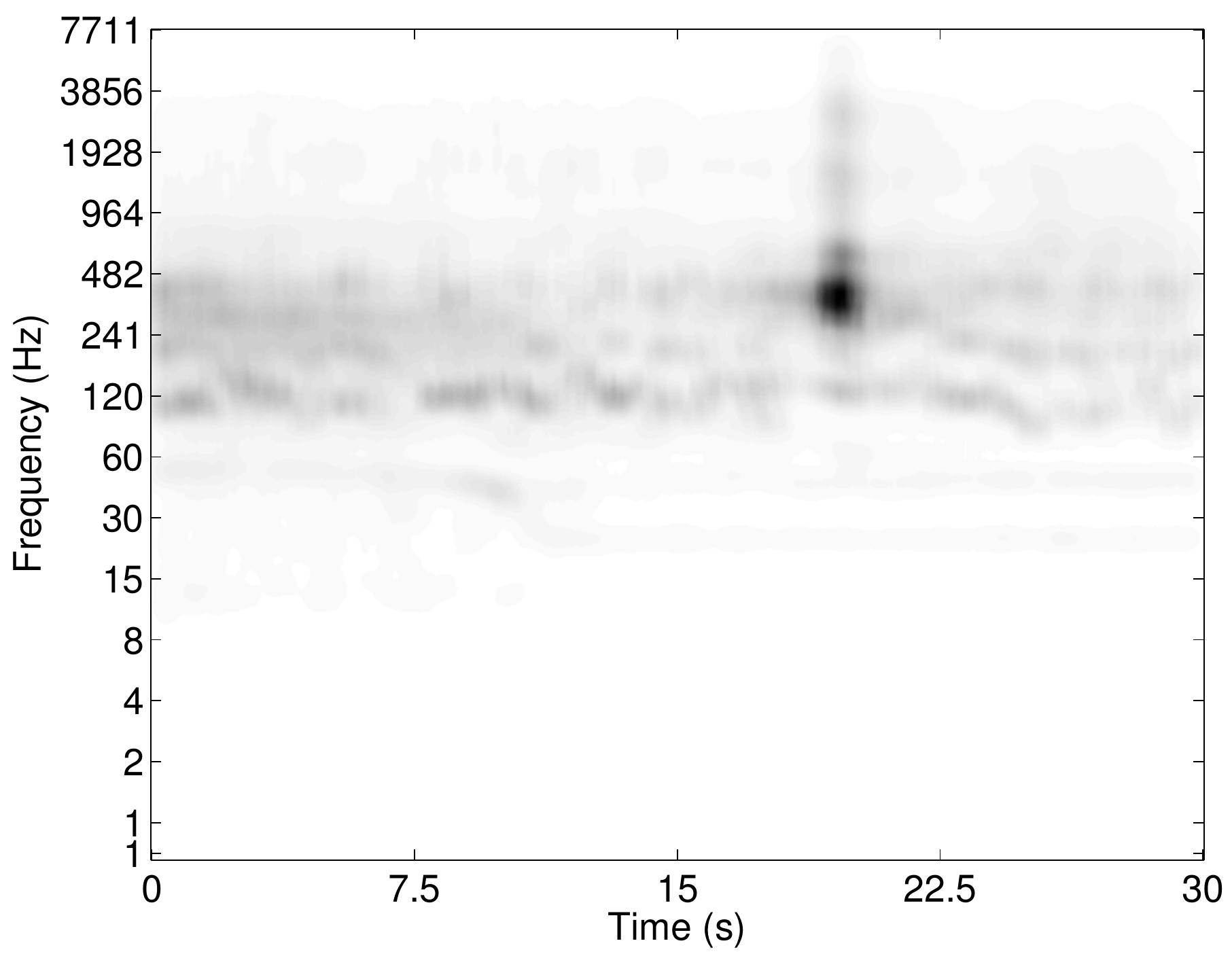}
~\hfill~ 
\includegraphics[width=3.5cm]{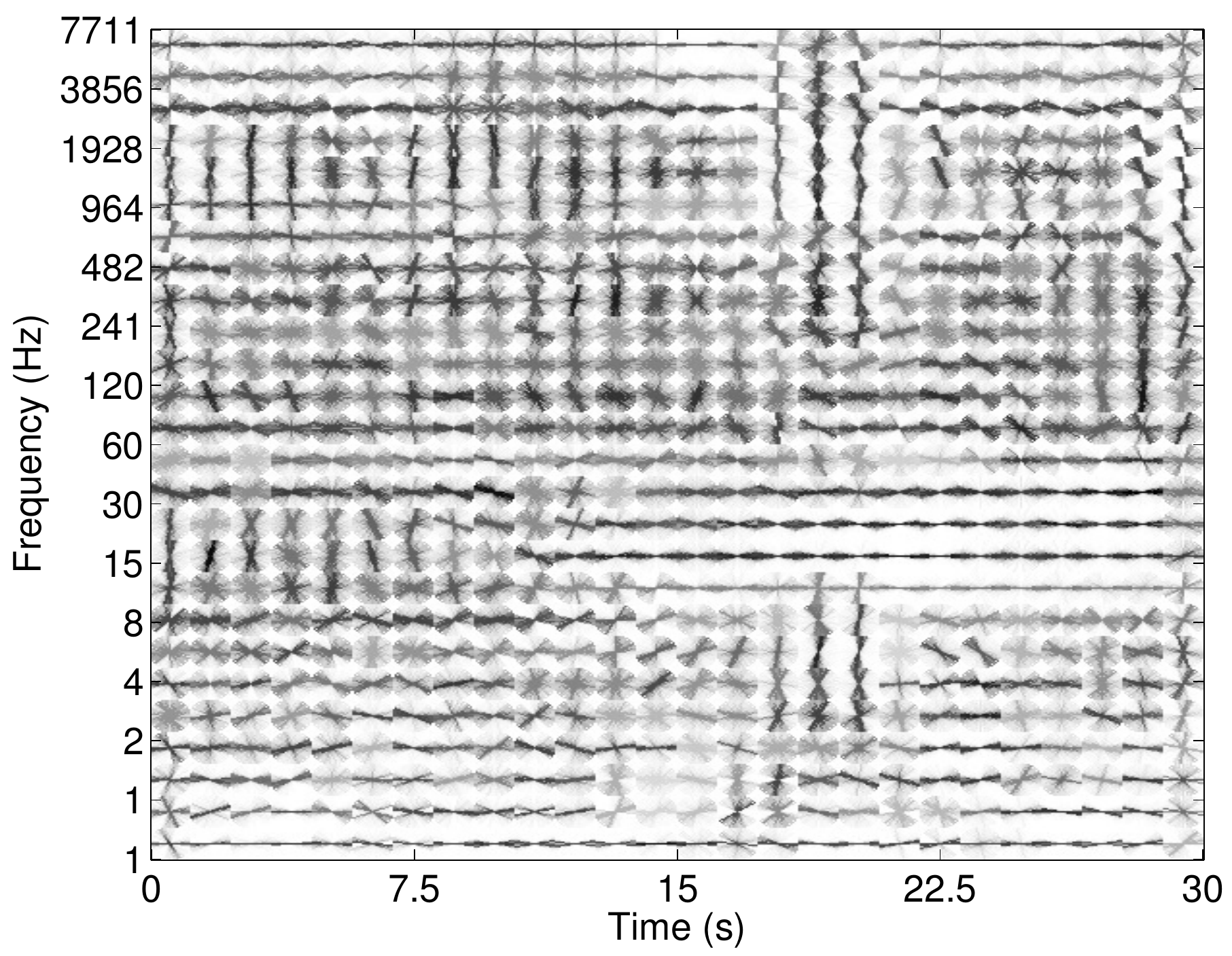}
\hfill~ \\
~\hfill (car) \hfill~\\
\vspace{0.3cm}
~\hfill  \includegraphics[width=3.5cm]{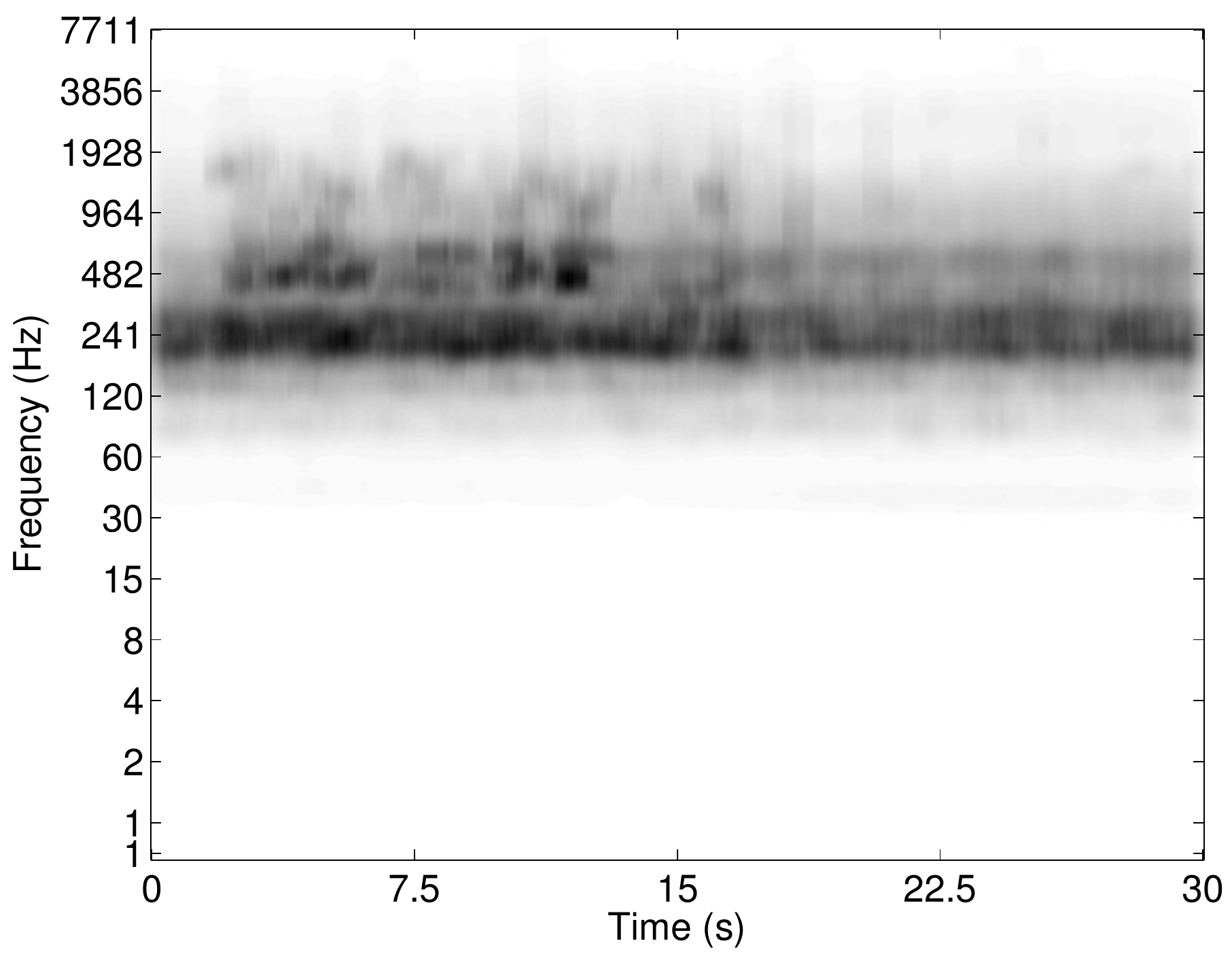}
~\hfill~ 
\includegraphics[width=3.5cm]{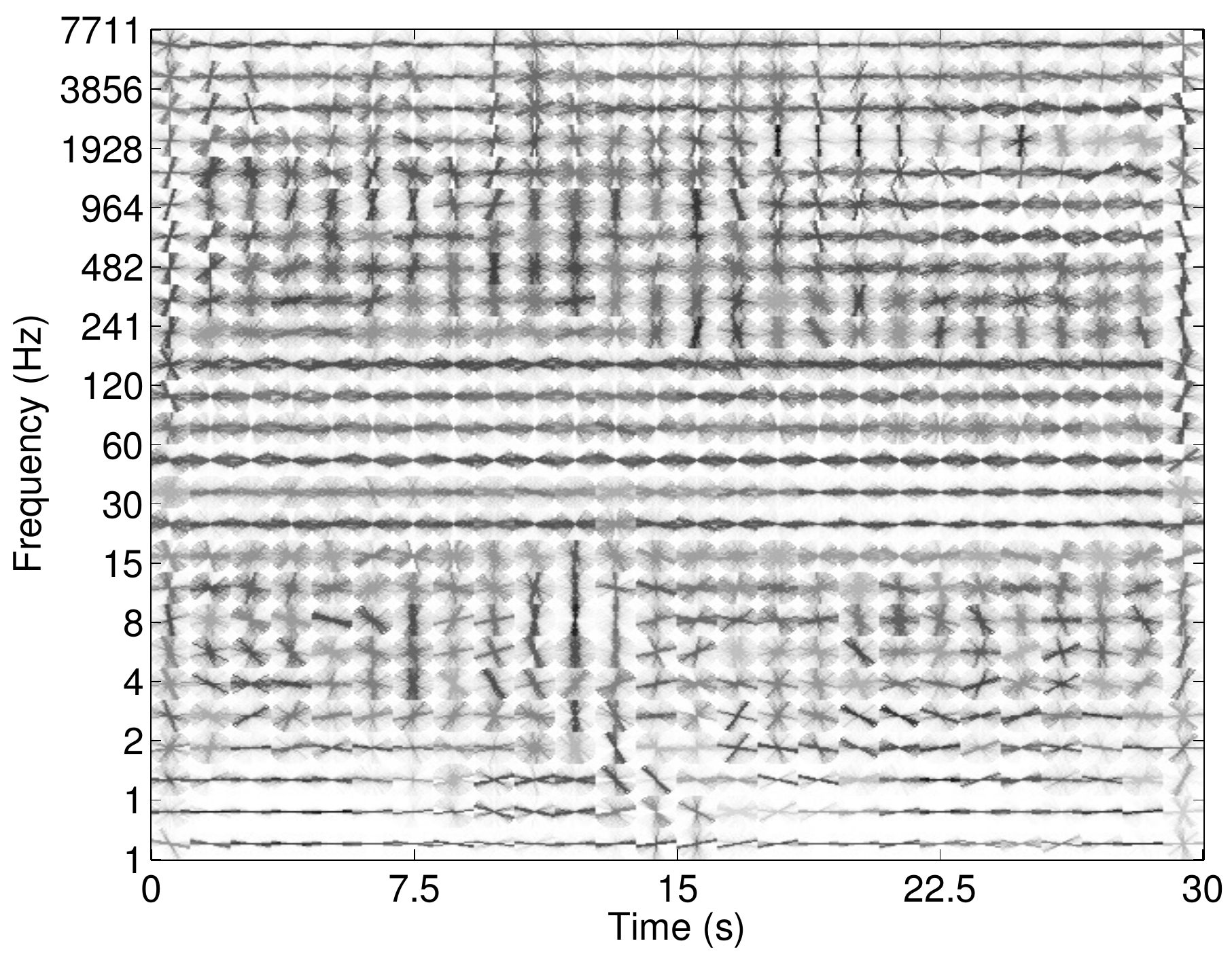}
\hfill~
\includegraphics[width=3.5cm]{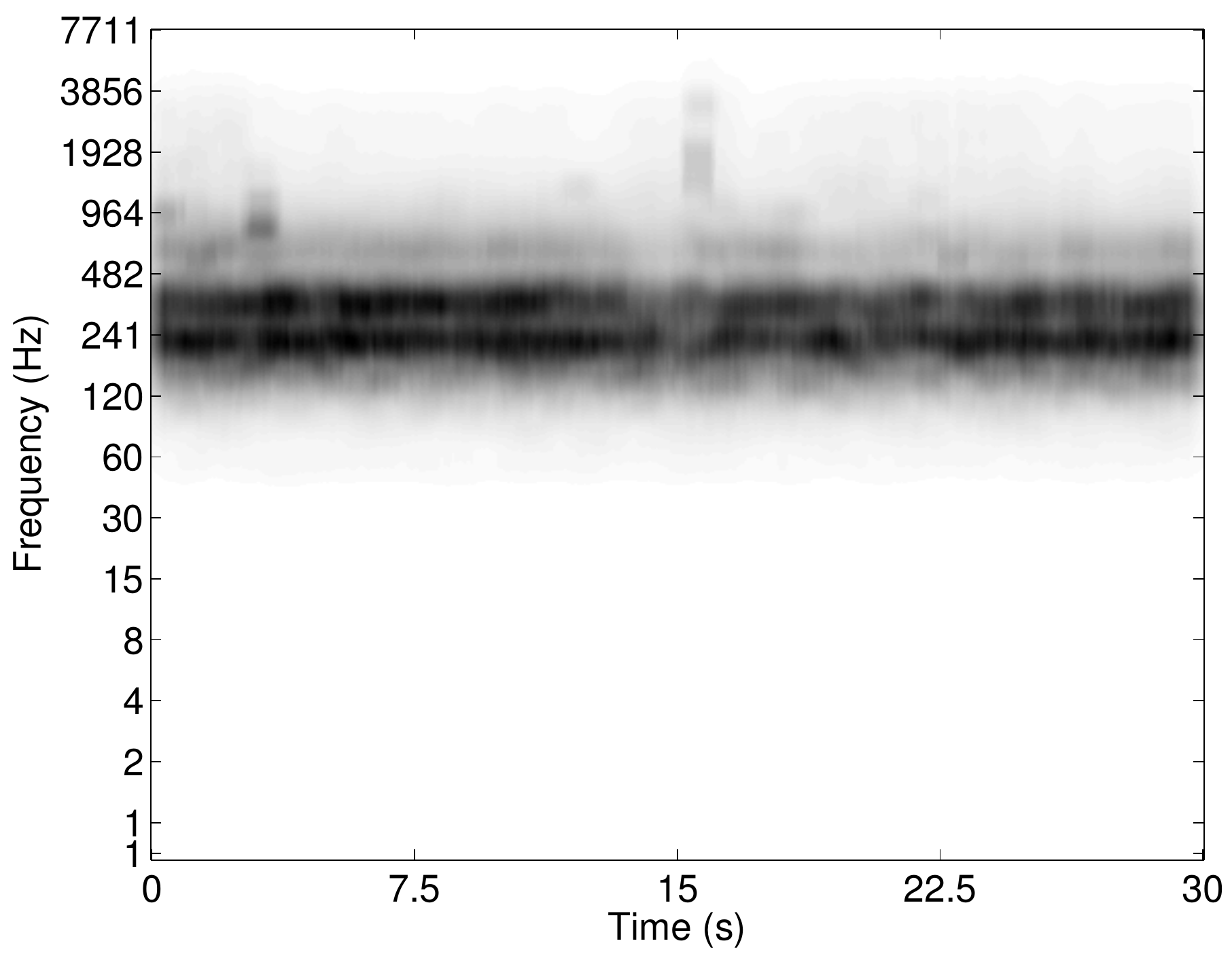}
~\hfill~ 
\includegraphics[width=3.5cm]{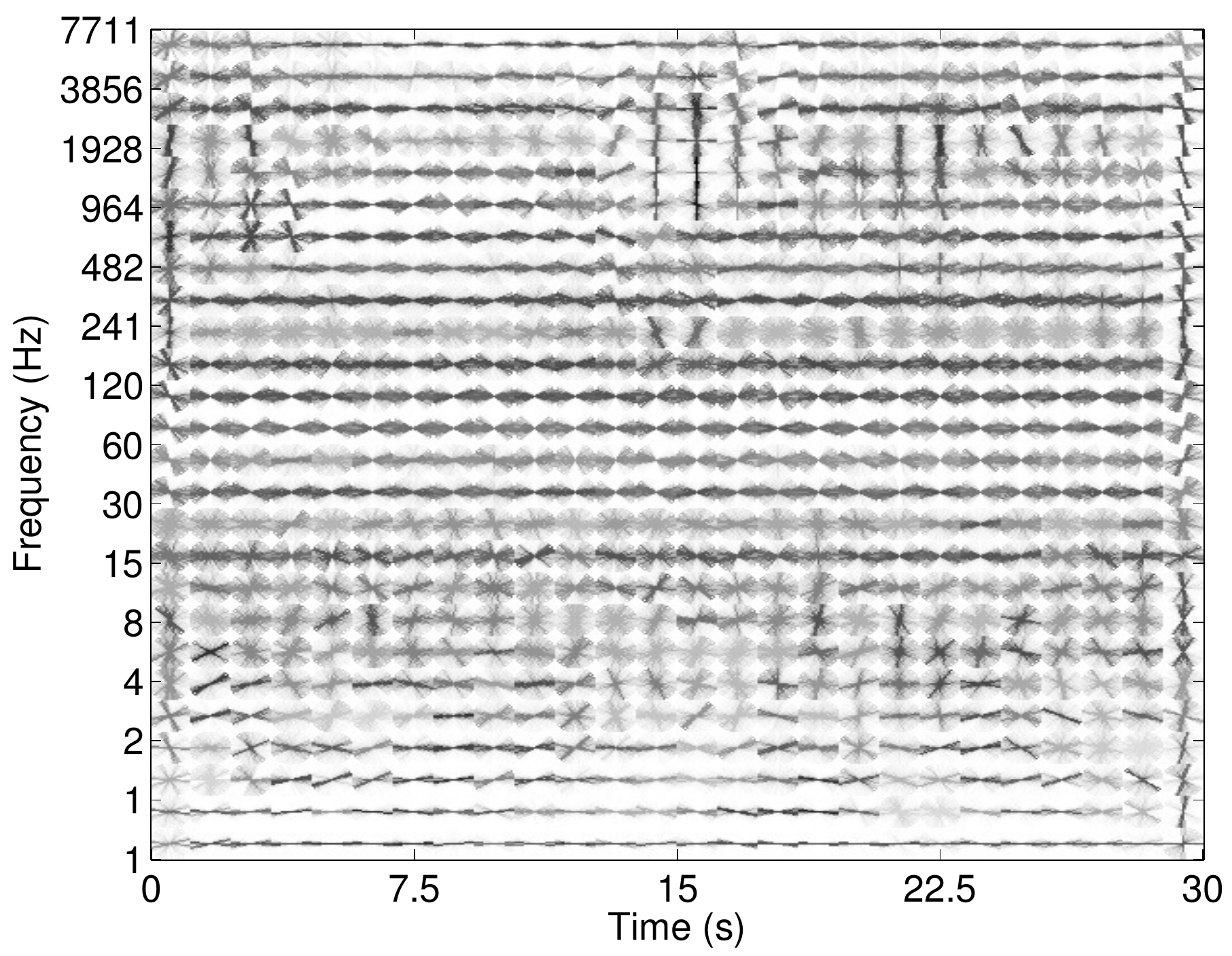}
\hfill~ \\
~\hfill (plane) \hfill~
\end{center}
  \caption{Examples of CQT representation and associated HOG representations
for transportation devices (top) bus. (middle) car. (bottom) plane.}
  \label{fig:hogtransport}
\end{figure*}

\begin{figure*}[t]
\begin{center}
~\hfill
  \includegraphics[width=3.5cm]{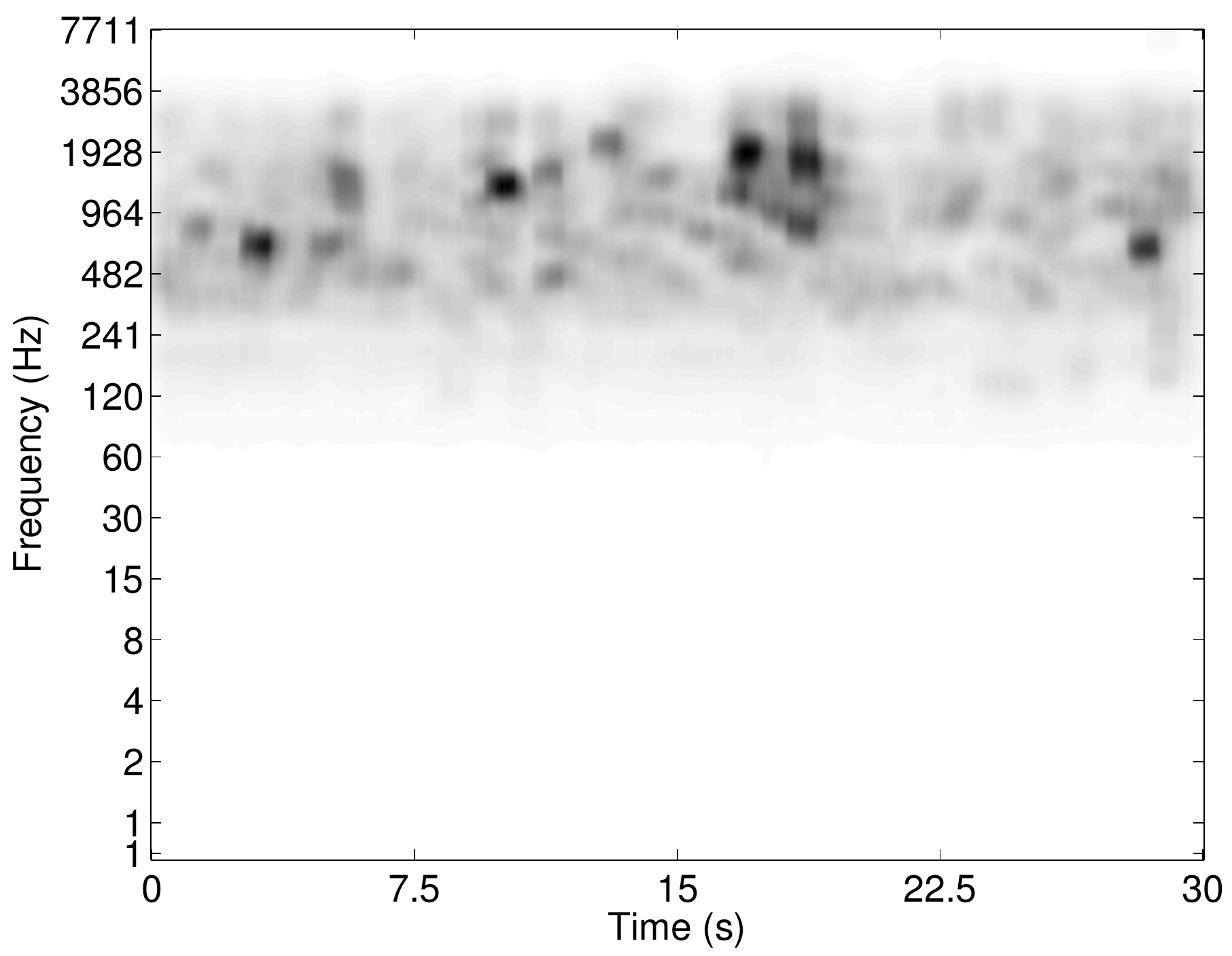}
~\hfill~ 
\includegraphics[width=3.5cm]{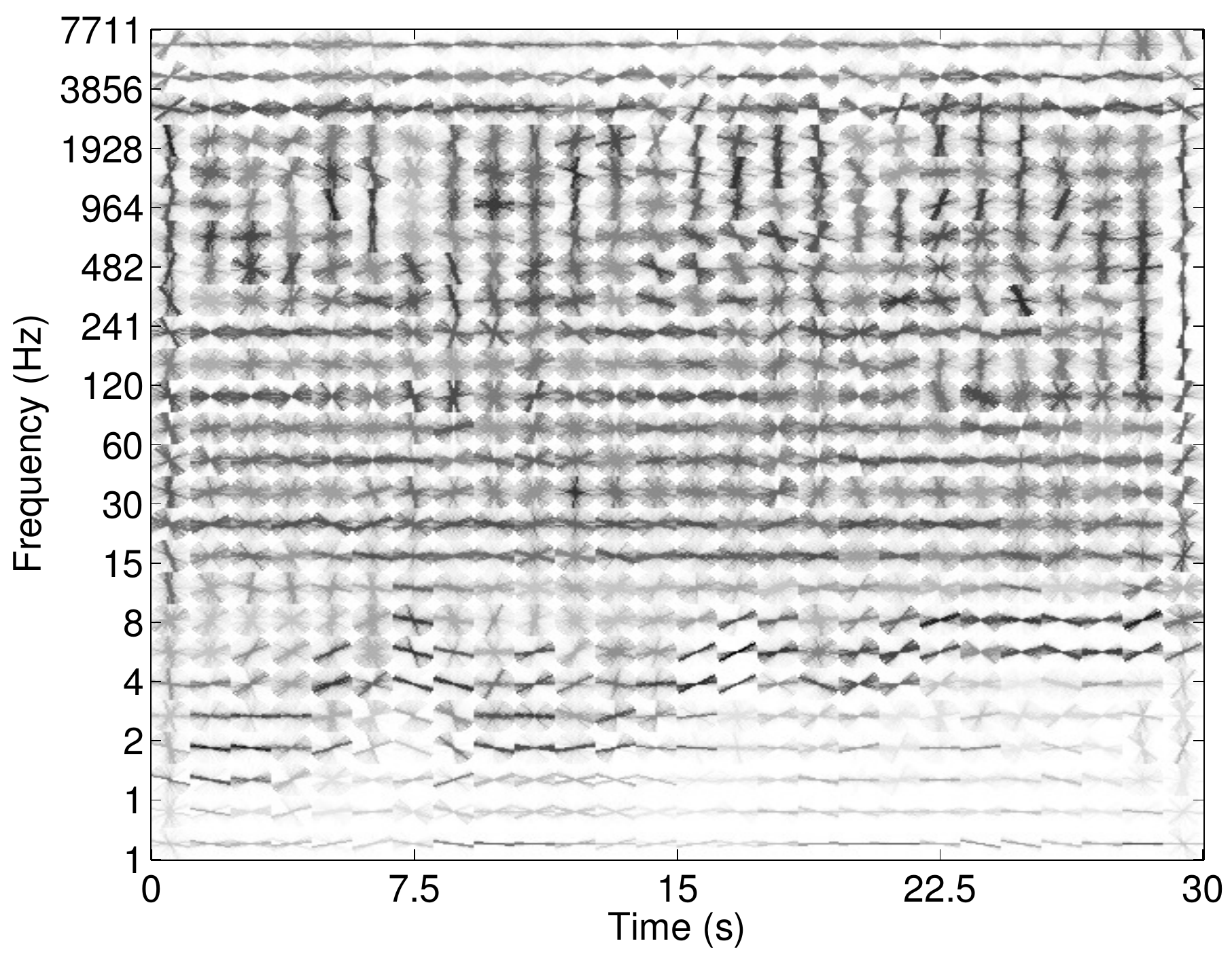}
\hfill~
\includegraphics[width=3.5cm]{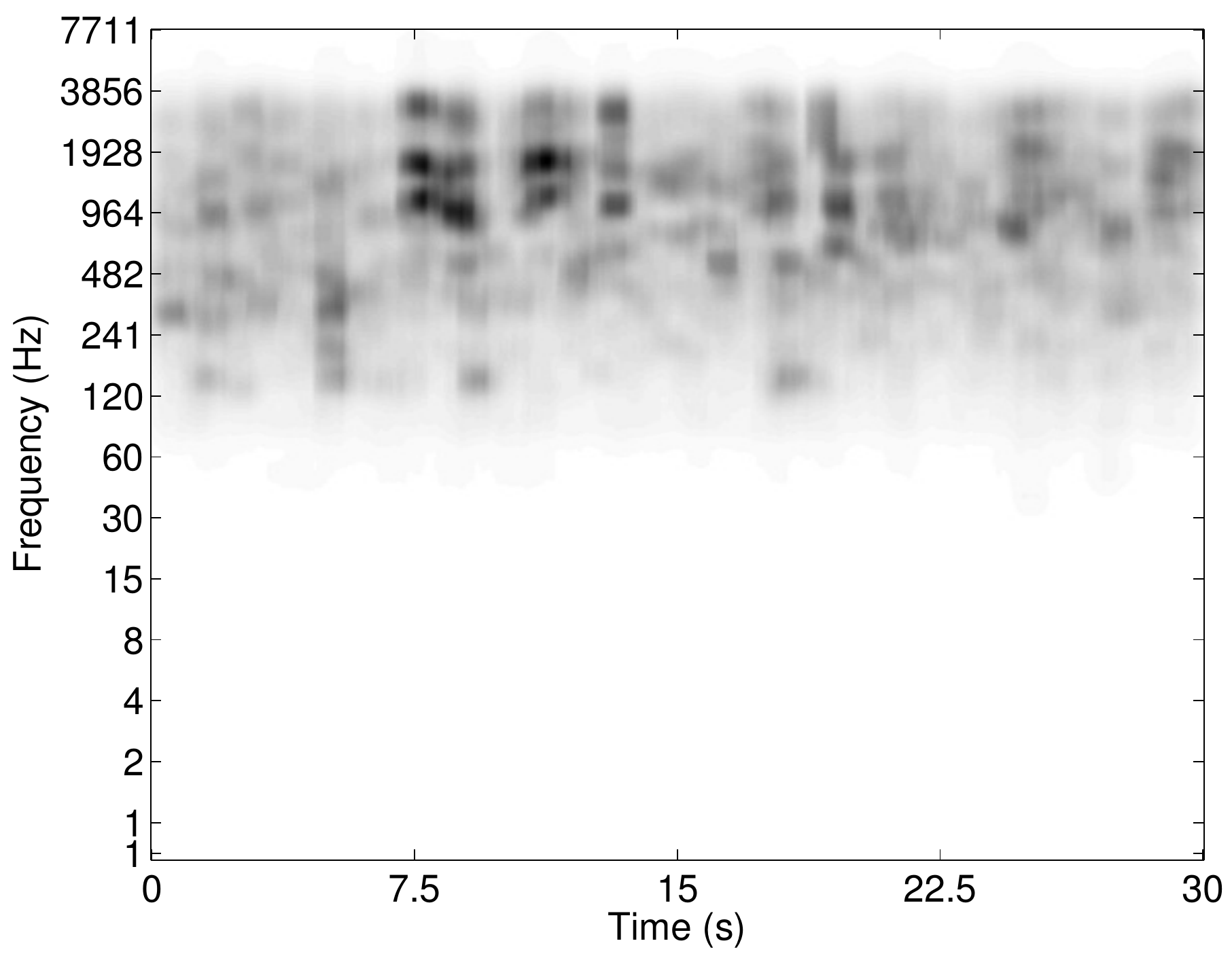}
~\hfill~ 
\includegraphics[width=3.5cm]{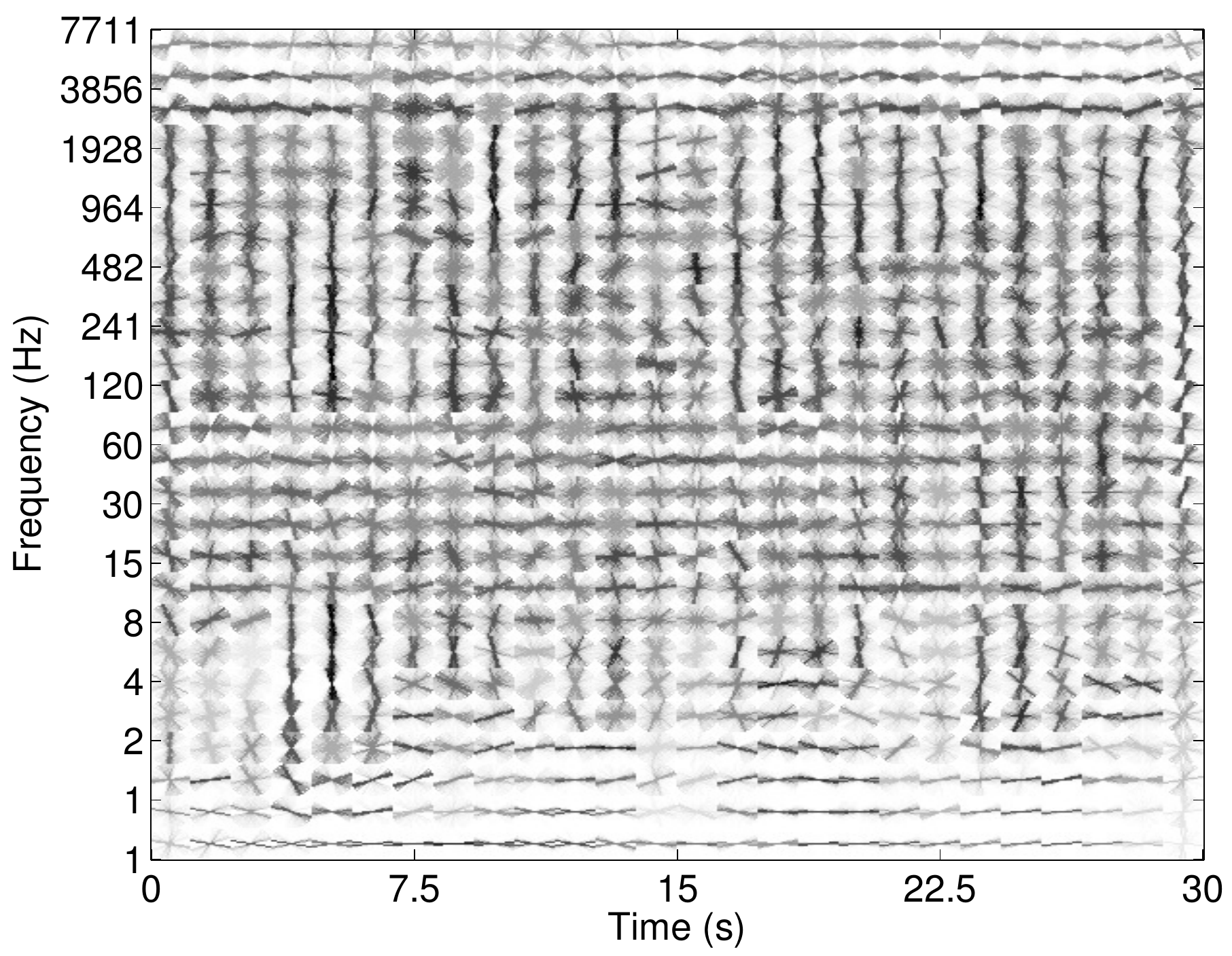}
\hfill~ \\
~\hfill (kid game hall) \hfill~ \\
\vspace{0.3cm}
~\hfill
  \includegraphics[width=3.5cm]{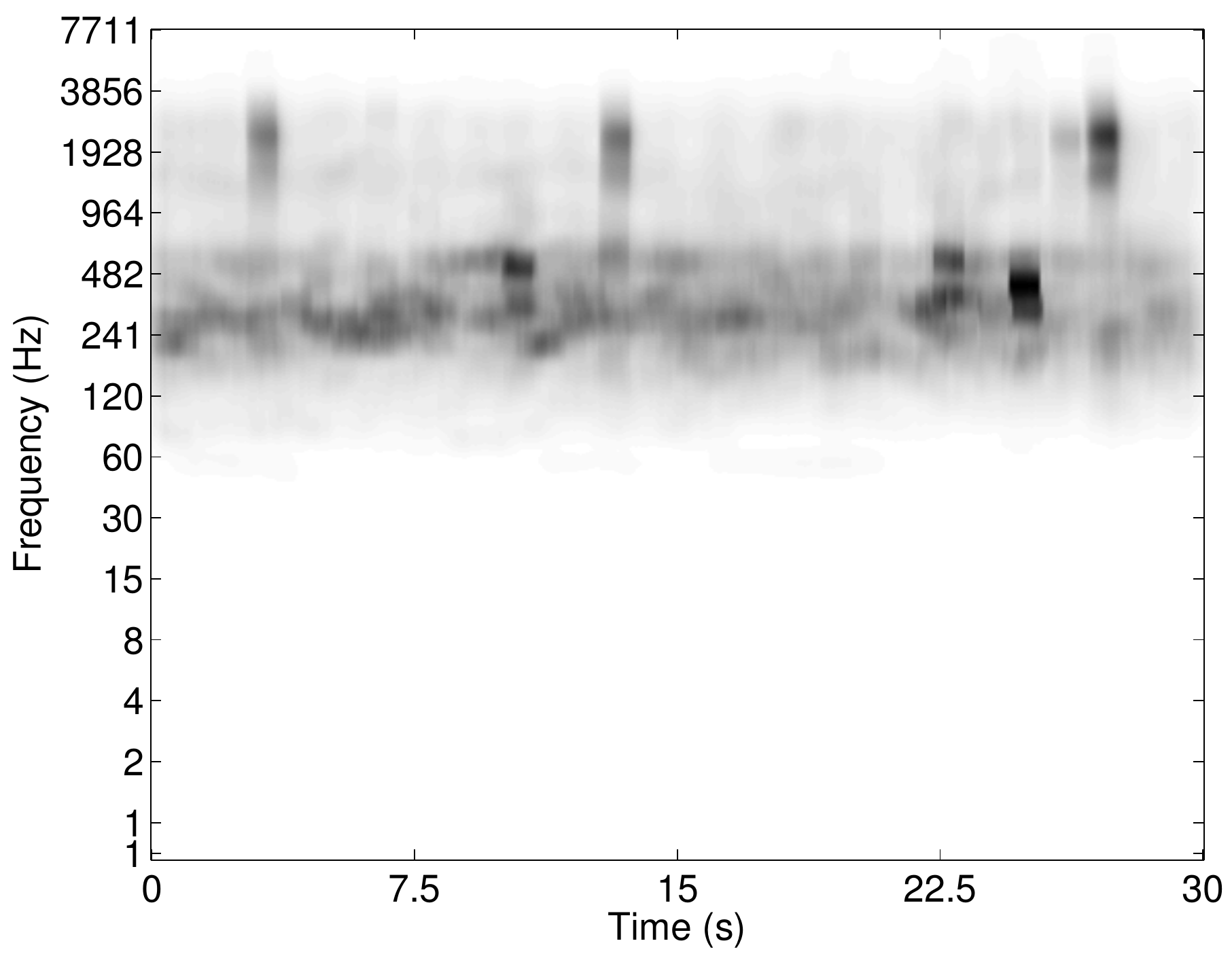}
~\hfill~ 
\includegraphics[width=3.5cm]{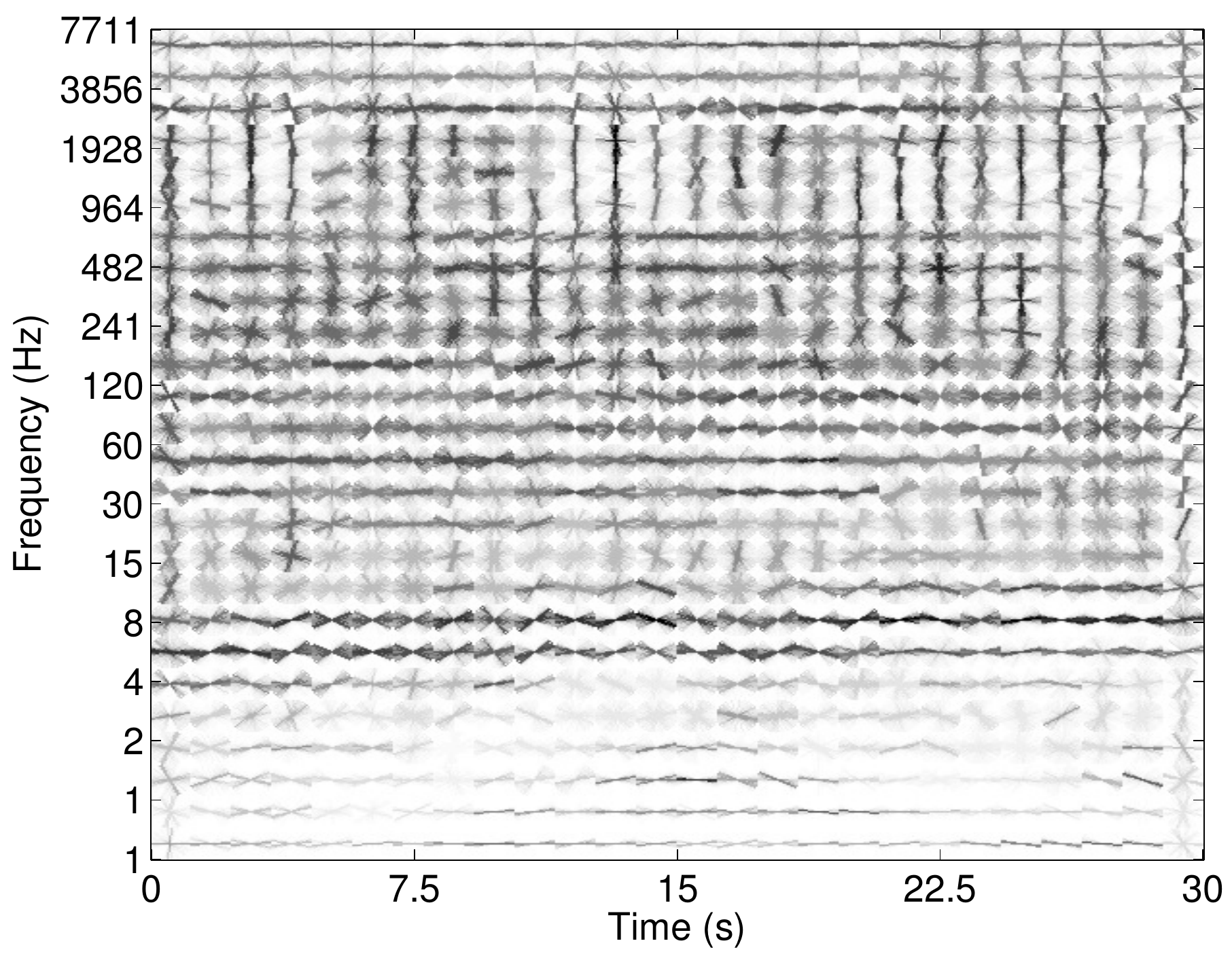}
\hfill~
\includegraphics[width=3.5cm]{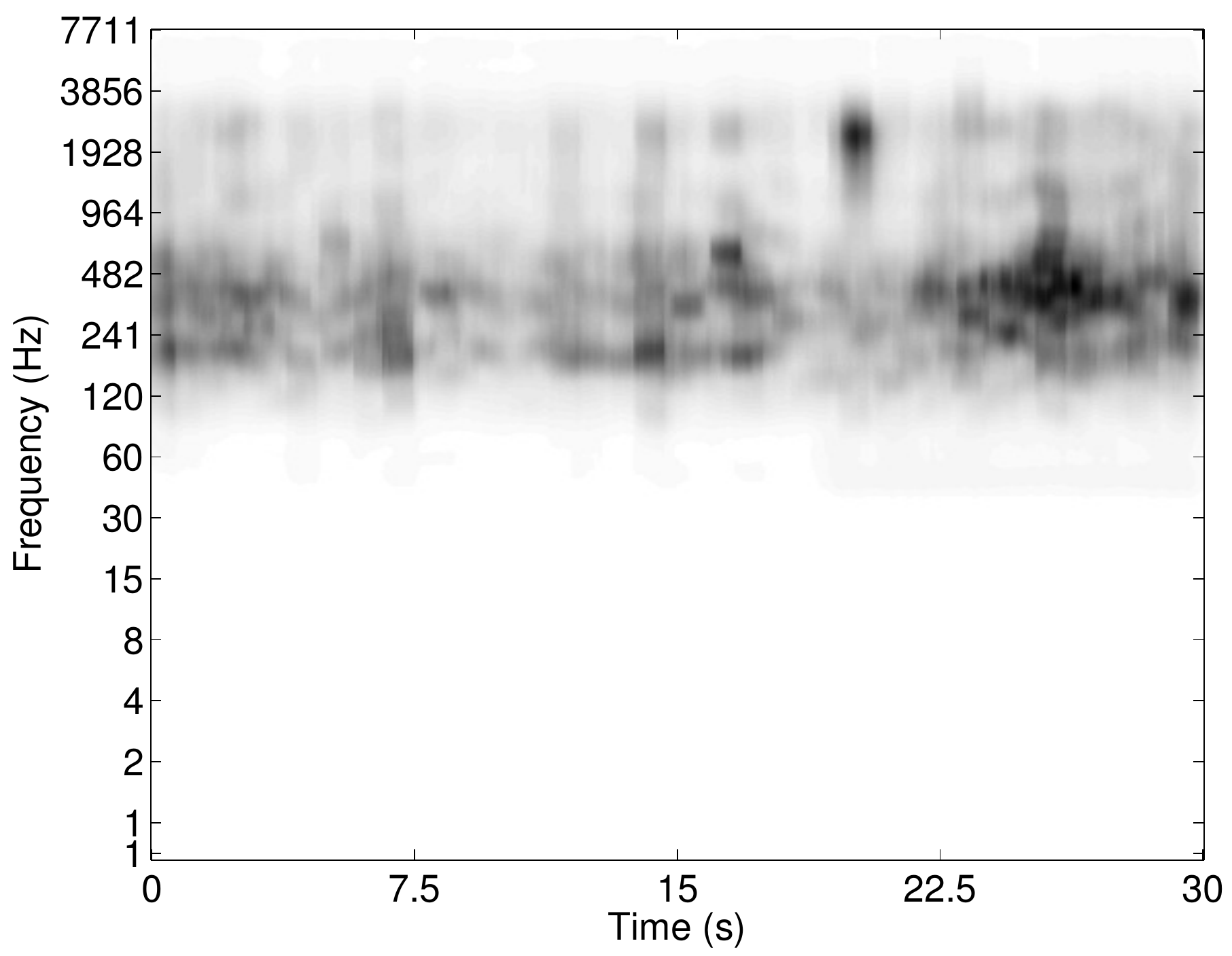}
~\hfill~ 
\includegraphics[width=3.5cm]{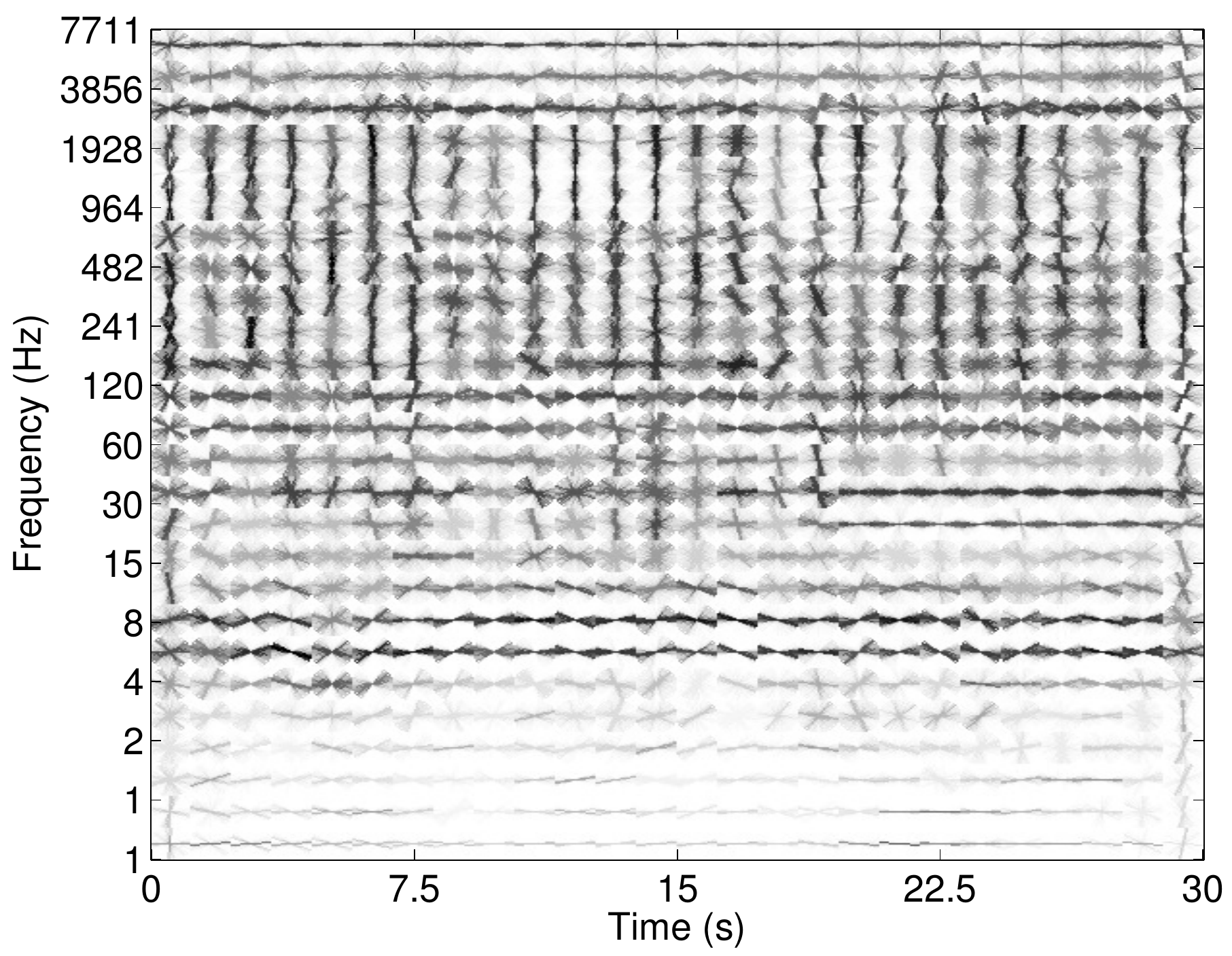}
\hfill~ \\
~\hfill (billiard pool hall) \hfill~\\
\end{center}
  \caption{Examples of CQT representation and associated HOG representations
for audio scene with babble noises and short-time events (top) kid game hall (bottom) pool hall.}
  \label{fig:babbleshort}
\end{figure*}

\modif{From Table \ref{tab:confmatrirouen},
we can first note a group of conveyances  \emph{plane}, \emph{bus}, \emph{car}, \emph{train} and \emph{high-speed train} that are precisely recognized (with precisions above $97\%$). Figure~\ref{fig:hogtransport} shows
examples of  CQT and  HOG representation of some audio scenes
related to these classes. Each class has specific signature
with time-frequency structures around $30$ Hz, $60$ Hz and $240$ Hz
for \emph{bus}, \emph{car} and \emph{plane}.
Two other classes of conveyance
 \emph{metro-rouen} and \emph{metro-paris} are sometimes mislabelled
showing that our HOG feature is not able to totally capture the fine-grained
discriminative features between these two classes {(if there is any)}.  

A  group of audio scenes composed of  speeches  (eventually loud ones) with specific short-time events (kid's scream and impacting balls) that occur all along the scenes is also precisely recognized with precision above $96\%$. 
These classes are \emph{kid game hall} and \emph{billiard pool hall}.
  Figure~\ref{fig:babbleshort} shows
examples of  CQT and  HOG representation of these audio scenes.
In these plots, the short time-frequency structures appearing around $2000$ Hz correspond to shouts and impacting balls.
 }

\begin{figure*}[t]
\begin{center}
~\hfill
  \includegraphics[width=3.5cm]{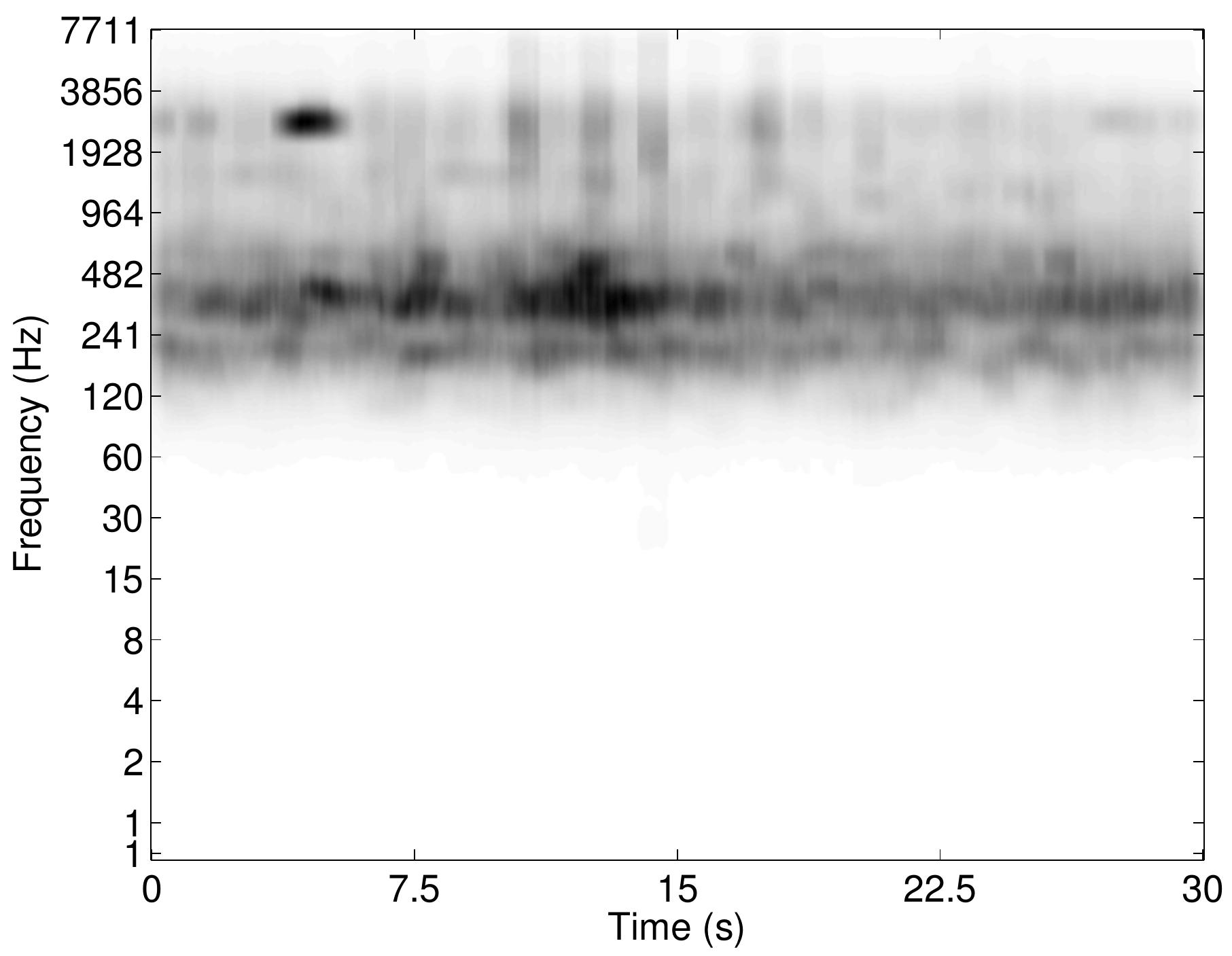}
~\hfill~ 
\includegraphics[width=3.5cm]{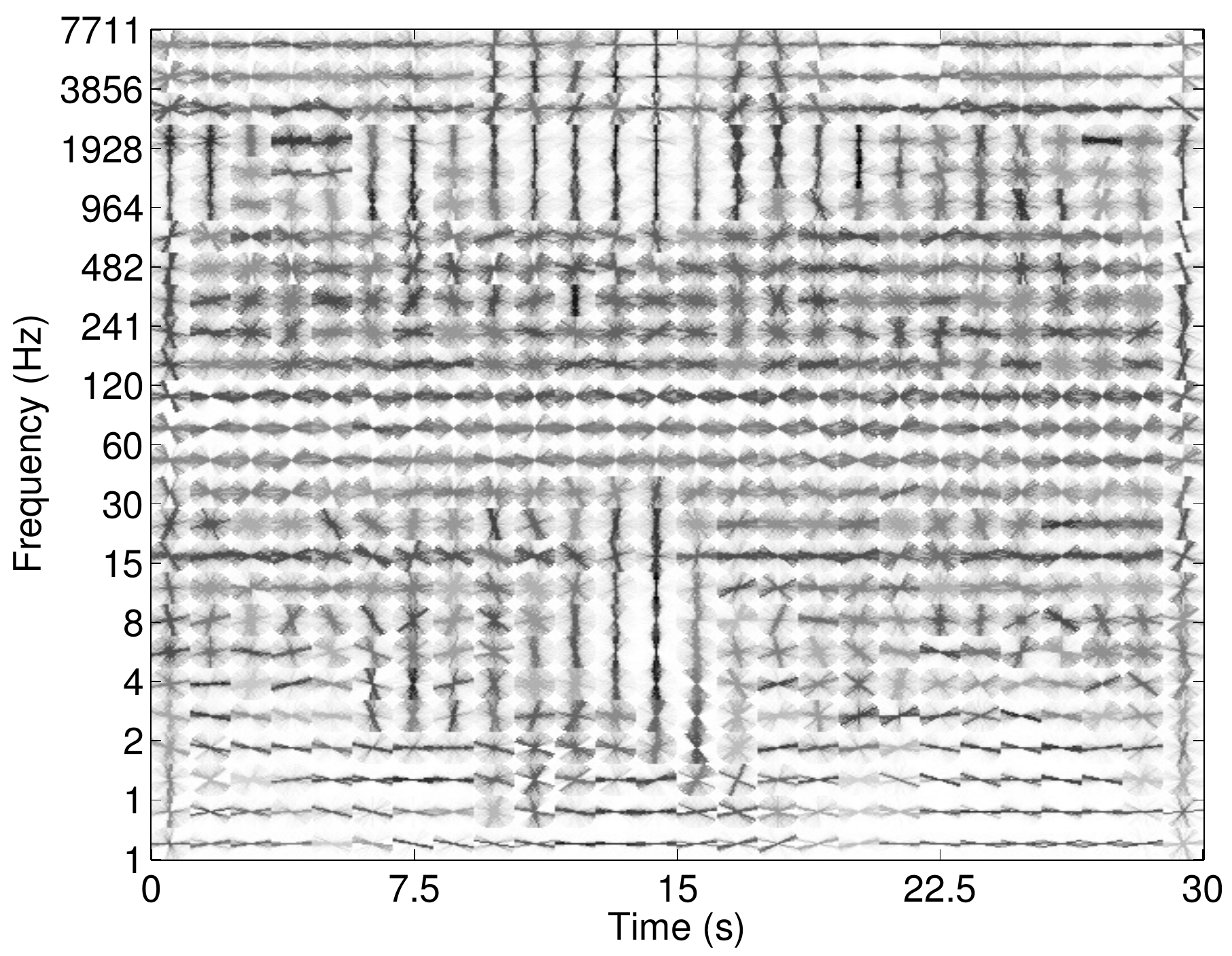}
\hfill~
\includegraphics[width=3.5cm]{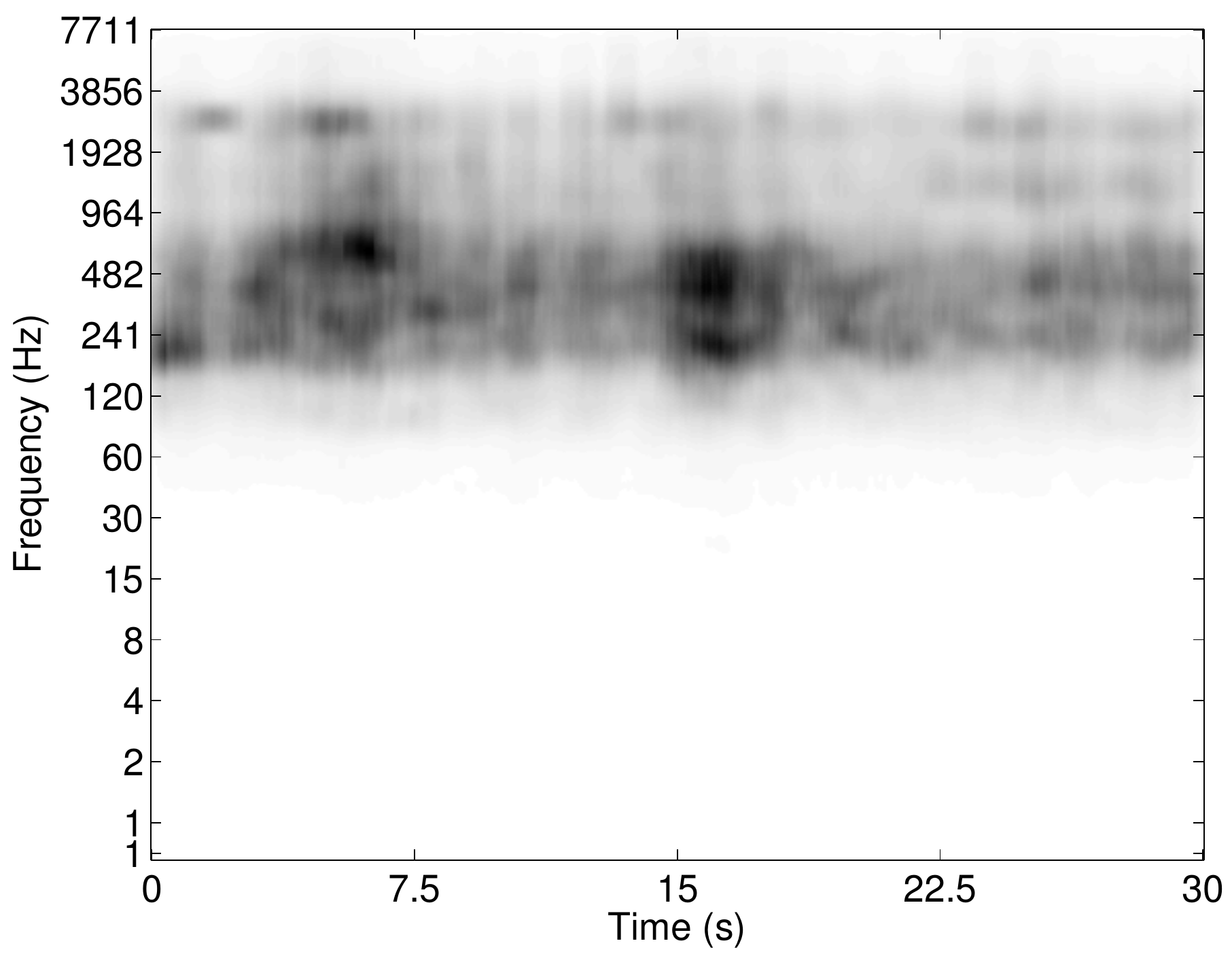}
~\hfill~ 
\includegraphics[width=3.5cm]{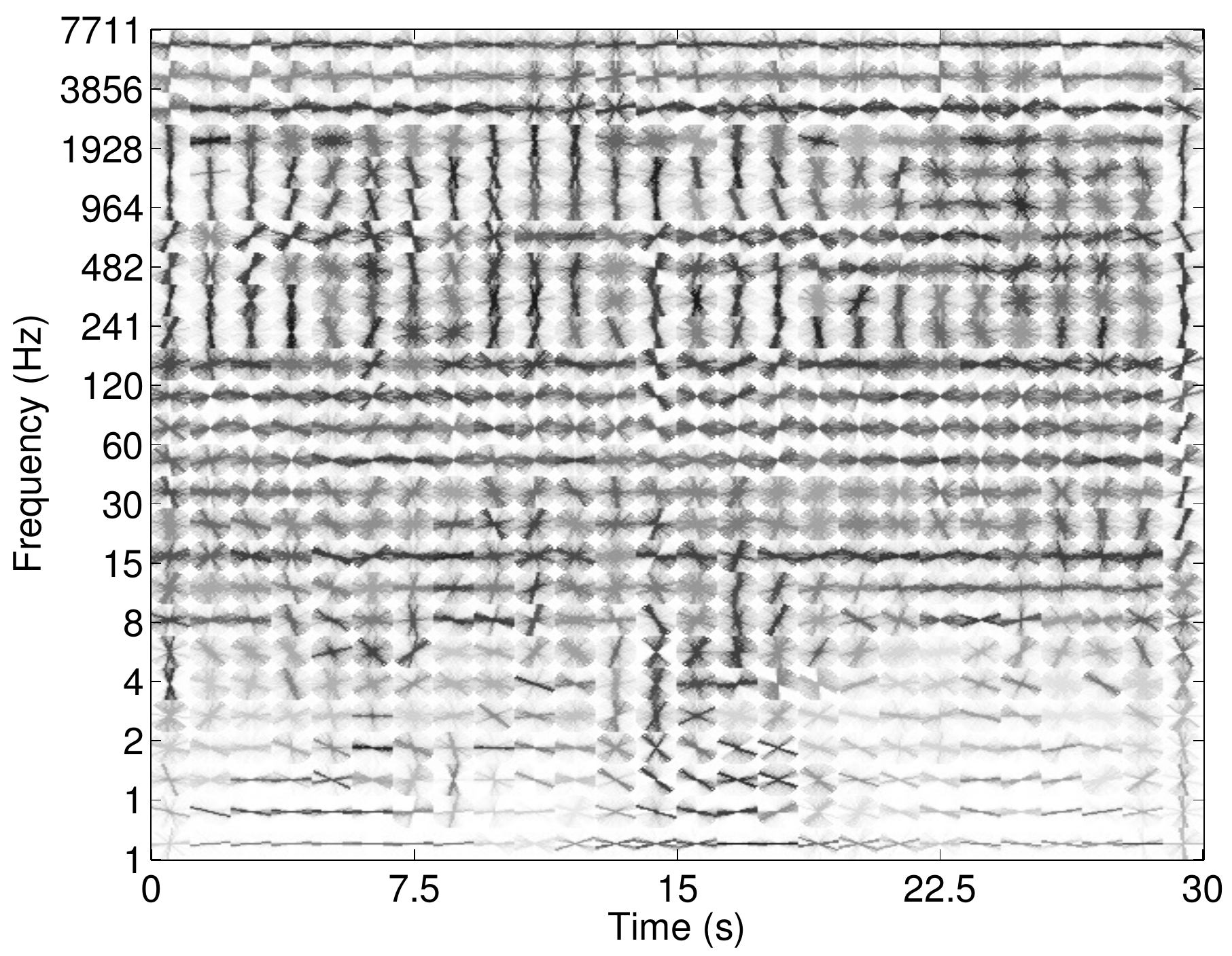}
\hfill~ \\
~\hfill (train station hall) \hfill~ \\
\vspace{0.3cm}
~\hfill
  \includegraphics[width=3.5cm]{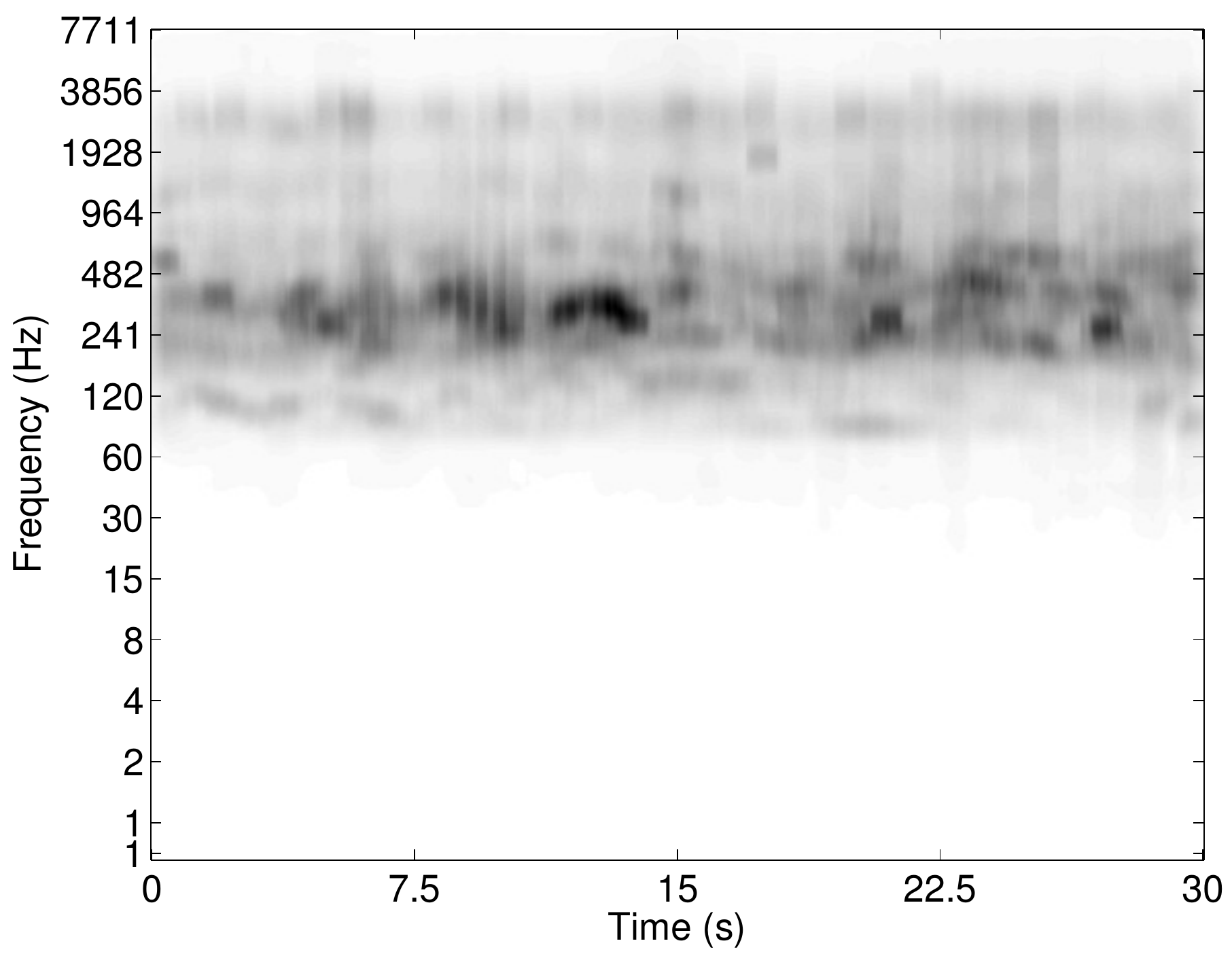}
~\hfill~ 
\includegraphics[width=3.5cm]{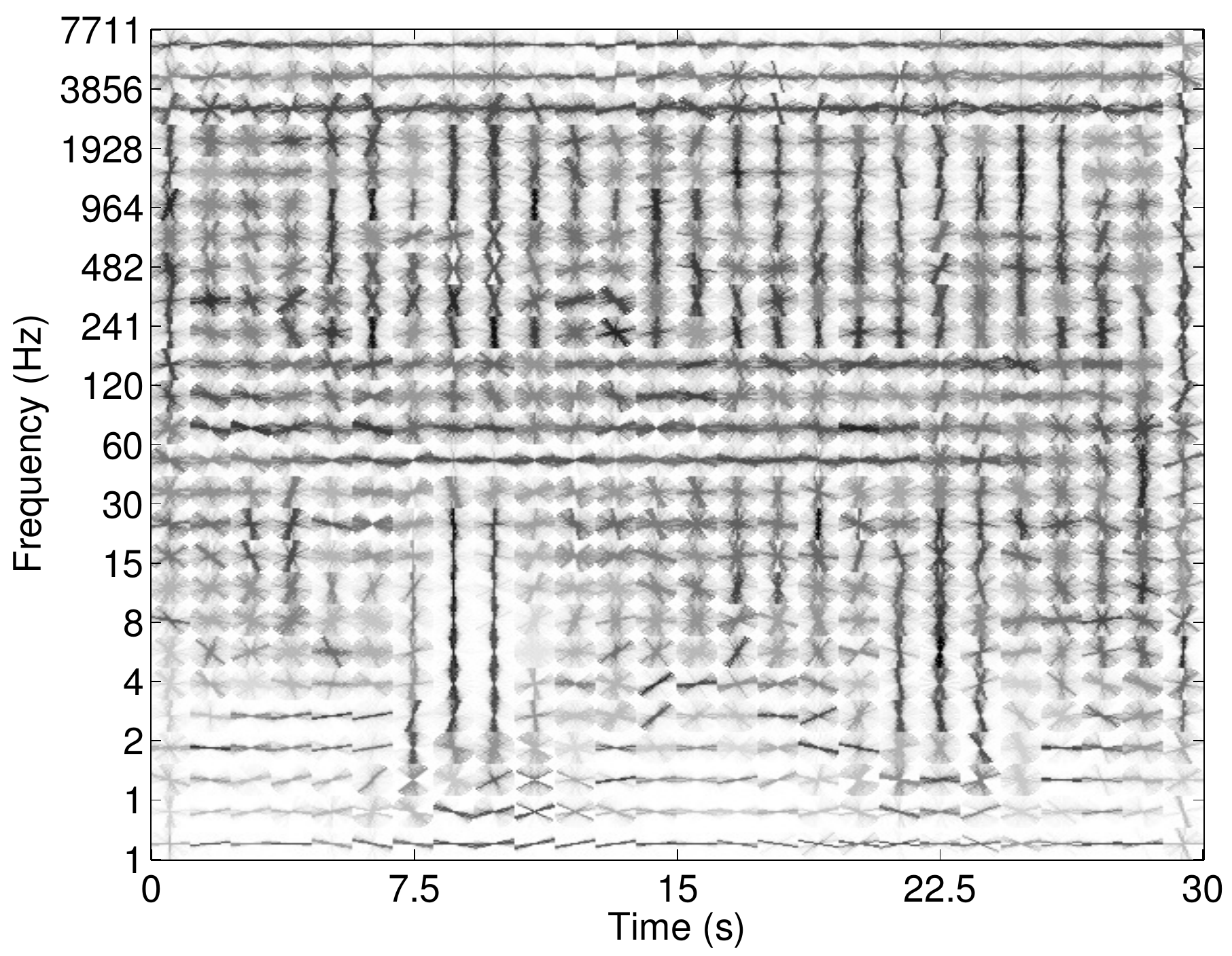}
\hfill~
\includegraphics[width=3.5cm]{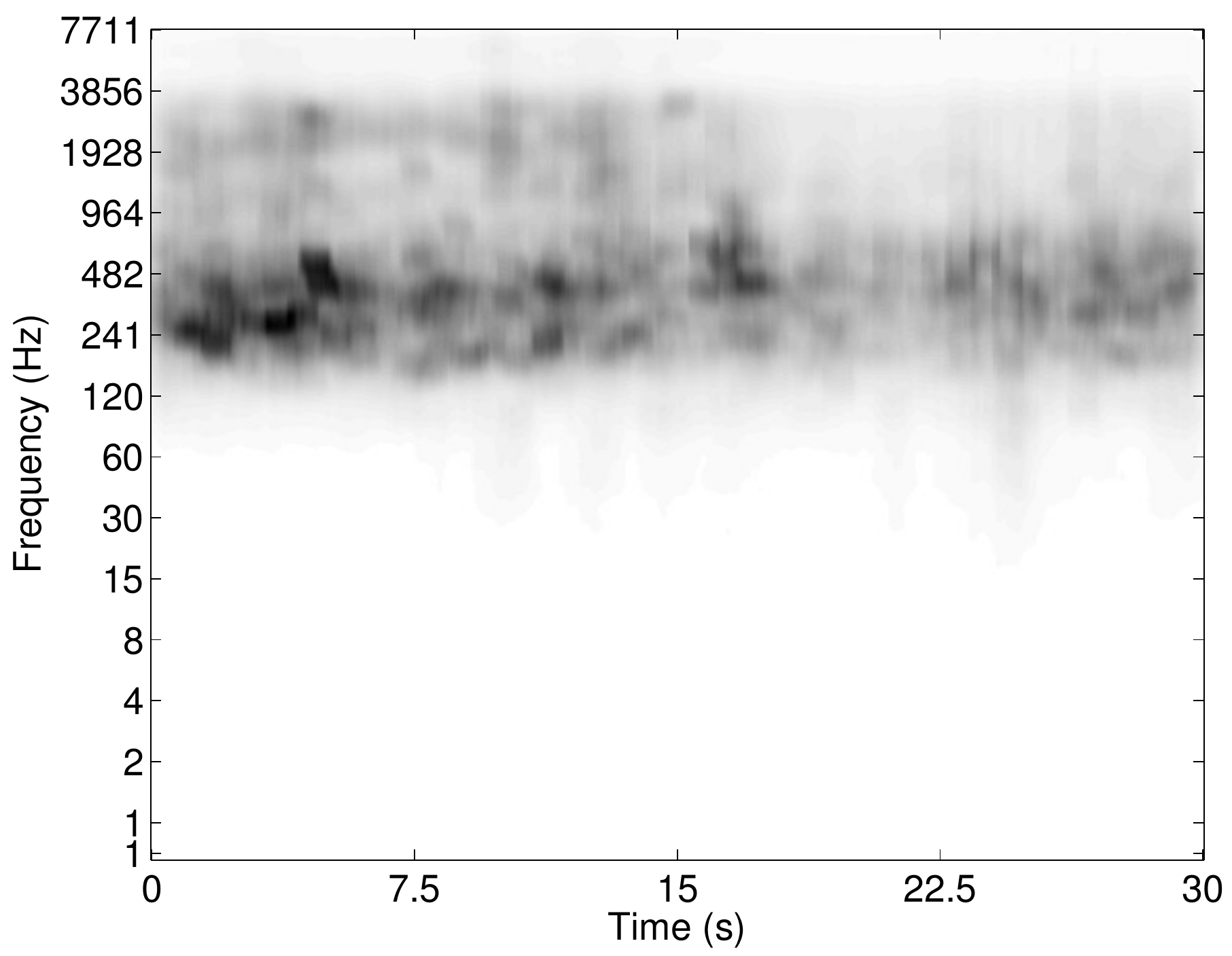}
~\hfill~ 
\includegraphics[width=3.5cm]{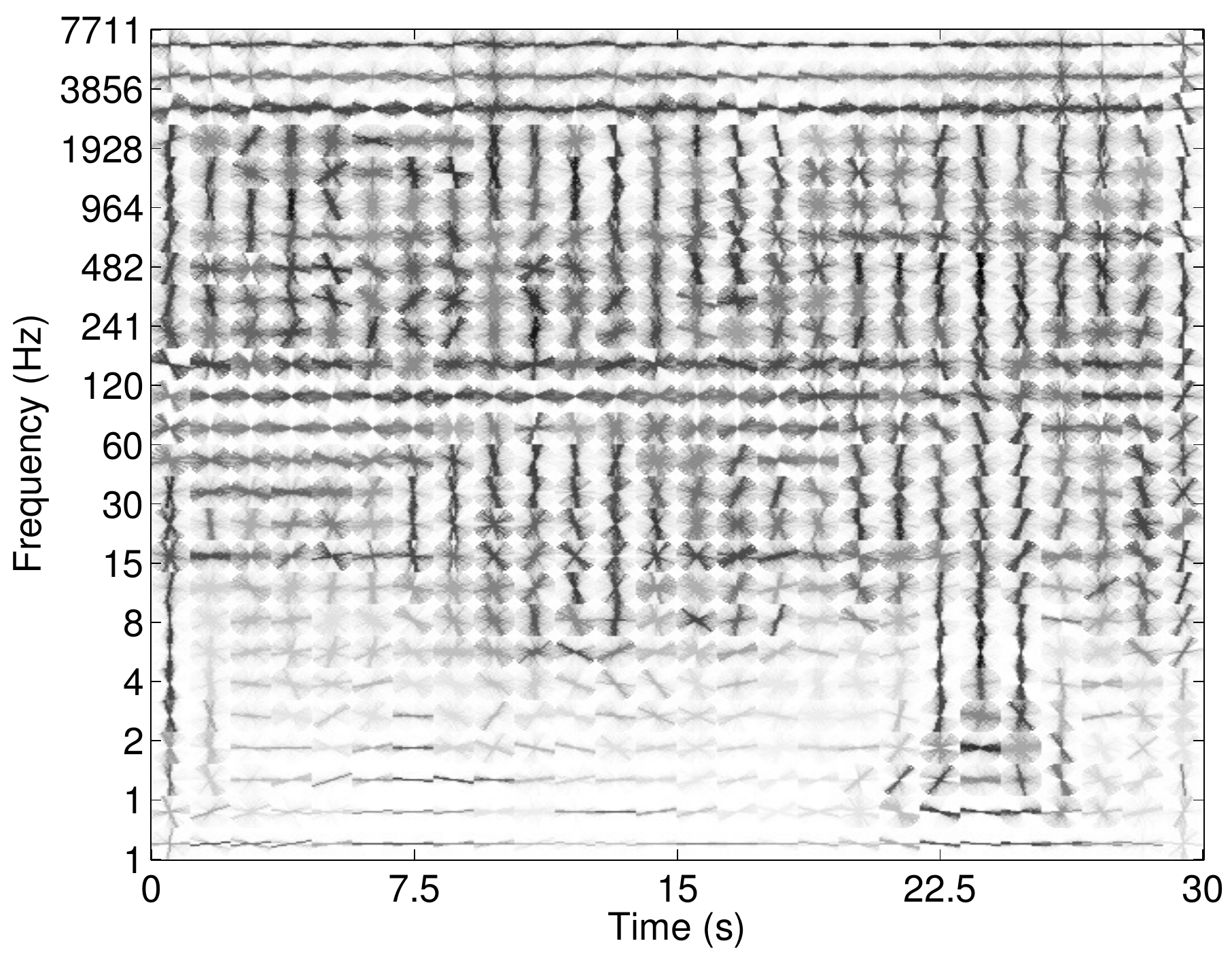}
\hfill~ \\
~\hfill (pedestrian street) \hfill~\\
\end{center}
  \caption{Examples of CQT representation and associated HOG representations
for audio scene with several people walking (top) train station hall (bottom) pedestrian street.}
  \label{fig:walkingpeople}
\end{figure*}

We can also notice a group of scenes that seems
difficult to distinguish as seen in Fig. \ref{fig:walkingpeople}: the ones in which some  people are walking, namely
\emph{pedestrian street}, \emph{market}, \emph{train station hall},
\emph{quiet street} and \emph{shop}.  These confusions are easily understandable. For instance, the main difference between a quiet
street and pedestrian street would be the number of people walking
in the scene. Such a difference is hardly taken into account by 
our HOG feature.

Despite the good skills exhibited by the proposed HOG features, the latter remarks show that 
there are still rooms for improvement, by
addressing the issues raised by classes that are  mixed. 
We believe that these classes show the needs for features (that
can still be based on HOG) capturing
discriminative short-time events that come in complement
to our ``global'' features.

\section{Conclusion}
The problem of classifying audio scene is currently a hot topic
in the computational auditory scene analysis domain. For this specific
problem, we have introduced in this paper a novel feature that seems
to be very promising at capturing relevant discriminative informations.
The main block of the feature we proposed has been initially proposed in the computer vision domain, namely histogram of gradients.

 Our novel feature has
been obtained by computing histogram of gradients of a constant Q-transform
followed by an appropriate pooling.
We have experimentally proved  that these histograms of gradients were useful for capturing specific characteristics present in a time-frequency representation that classical features such as MFCC can not encode namely the local
variation of power spectrum.
Then, our experimental  results on real datasets clearly {show}
that our features achieve state-of-the-art classification performances
on several datasets. 

While our HOG-based feature is globally efficient, the overall pipeline 
for audio scene classification still lacks in discriminating some difficult classes. In order to further improve the scheme, some efforts are still needed. 
Our future researches focus on improving {discriminative} ability of HOG-based feature by working on the pooling strategy. The supervised learning paradigm
may also be improved by taking into account an hierarchical taxonomy of
the classes. We plan to {investigate} this taxonomy by learning it
directly from the data.

\bibliographystyle{IEEEtran}

\end{document}